\documentclass[aps,showpacs,twocolumn,superscriptaddress]{revtex4-2}
\usepackage[T1]{fontenc}
\usepackage{graphicx}% Include figure files
\usepackage{dcolumn}% Align table columns on decimal point
\usepackage{bm}% bold math
\usepackage{booktabs}
\usepackage{color}
\usepackage[normalem]{ulem} % \sout{old text} for strikeout
\usepackage[dvipsnames]{xcolor} % For blue in-text comments
\usepackage{hyperref}
\hypersetup{
  colorlinks=true,        % false: boxed links; true: colored links
  linkcolor=blue,         % color of internal links
  citecolor=cyan,         % color of links to bibliography
}
\usepackage{mathrsfs}
\usepackage[utf8]{inputenc}
\usepackage{mathtools}

\usepackage{doi}
\usepackage{amsmath}
\usepackage{doi}
\usepackage{amssymb}

\begin{document}
	
%%%
\title{Low-mass X-ray binaries as a probe of Kerr-MOG black hole spacetime}
%\title{Kerr-MOG black hole: radiative efficiency and astrophysical relativistic jets}
%%%%

\author{Bakhodirkhon Saidov}
    \email[]{saidbaxodirxon13@gmail.com}

    \affiliation{Institute of Fundamental and Applied Research, National Research University TIIAME, Kori Niyoziy 39, Tashkent 100000, Uzbekistan}
    
    \affiliation{Institute for Advanced Studies, New Uzbekistan University, Movarounnahr str. 1, Tashkent 100000, Uzbekistan}
    
	%%%%%%%%%%
	\author{Bakhtiyor Narzilloev}
	\email{b.narzilloev@newuu.uz}
    \affiliation{Institute for Advanced Studies, New Uzbekistan University, Movarounnahr str. 1, Tashkent 100000, Uzbekistan}
 %%%%%%

 \author{Ibrar Hussain}
	\email{ibrar.hussain@seecs.nust.edu.pk}	
	\affiliation{School of Electrical Engineering and Computer Science, National University of Sciences and Technology, H-12, Islamabad, Pakistan}
    \affiliation{Research Center of Astrophysics and Cosmology, Khazar University, 41 Mehseti Street, AZ1096 Baku, Azerbaijan}
%%%%%%%%%%%
   
\author{Bobomurat Ahmedov}%,\orcidlink{0000-0002-1232-610X}}
\email{ahmedov@astrin.uz}

\affiliation{School of Physics, Harbin Institute of Technology, Harbin 150001, People’s Republic of China}

\affiliation{Institute of Theoretical Physics, National University of Uzbekistan, Tashkent 100174, Uzbekistan}
    
%\affiliation{Institute for Advanced Studies, New Uzbekistan University, Movarounnahr str. 1, Tashkent 100000, Uzbekistan}

\date{\today}
\begin{abstract}
We investigate the astrophysical implications of rotating black holes in modified gravity by studying the Kerr-MOG spacetime and applying it to several stellar-mass black hole candidates: A0620-00, H1743-322, XTE J1550-564, GRS1124-683, GRO J1655-40, and GRS1915+105. The Kerr-MOG geometry is characterized by the black hole mass, the spin parameter $a$, and the modified gravity parameter $\alpha$. We analyze how the presence of the MOG parameter modifies the spacetime structure and the location of the innermost stable circular orbit (ISCO). Within the Novikov--Thorne accretion disk model, we show that $\alpha$ significantly affects the radiative efficiency of accretion disks and introduces a degeneracy between the spin and the modified gravity parameter. Using observational estimates of radiative efficiencies inferred from the continuum-fitting method, we constrain the allowed regions in the $(\alpha,a/M)$ parameter space for each source. We further examine the relativistic jet power using the Blandford--Znajek mechanism and compare the theoretical predictions with the observed transient jet energetics, considering two representative jet Lorentz factors, $\Gamma=2$ and $\Gamma=5$. By combining the constraints from the radiative efficiency and jet power, we identify regions where both observables can be simultaneously reproduced. For several sources significant overlap regions appear, while for the highly spinning source GRS1915+105 the compatibility occurs only within a narrow range of Kerr--MOG parameters. These results suggest that the Kerr--MOG spacetime can provide a viable framework for interpreting the observed properties of several black hole X-ray binaries.
\end{abstract}

\maketitle

\section{Introduction}

Modifications of General Relativity (GR) are widely regarded as necessary to address several fundamental challenges, including the accelerated expansion of our Universe \cite{mog1}, the issue of spacetime singularities \cite{mog2}, and the quantization of the gravitational field on a curved background \cite{mog3}, which remain unresolved within its theoretical framework. In this context, various modified or alternative theories have been proposed to overcome these issues (see, for example, \cite{mog4,mog5,mog6}). The parameters of these modified or alternative theories of gravity have been constrained by studying particle dynamics in black hole spacetimes derived within these frameworks, as well as related phenomena such as black hole shadows and gravitational lensing \cite{mog7,mog8,mog9,mog10}, quasi-periodic oscillations \cite{mog11,mog12}, and quasinormal modes \cite{mog13,mog14}, through comparisons with observational data released by the Event Horizon Telescope collaboration \cite{mog15} and LIGO–Virgo \cite{mog16}. 

John Moffat \cite{mog17} proposed a modified theory of gravity (MOG) as a relativistic alternative to standard gravity, formulated within the framework of scalar–tensor–vector gravity (STVG) \cite{mog18}. The MOG theory was developed to address several outstanding astrophysical and cosmological problems, including galaxy rotation curves, the dark matter problem, cosmic acceleration, and large-scale structure formation, without invoking non-baryonic dark matter particles \cite{mog18,mog19}. In MOG, the Newtonian gravitational constant, the vector field coupling constant and the vector field mass are promoted to dynamical scalar fields. This generalization allows these effective constants to vary with spacetime, leading to a modified gravitational interaction that can account for enhanced gravitational effects at galactic and cosmological scales while remaining consistent with local gravitational tests.

Black hole solutions constitute a fundamental aspect of any gravitational theory, whether it is GR or a modified or alternative theory of gravity. Consequently, they have been extensively studied in a wide range of contexts, including particle dynamics and black hole thermodynamics \cite{mog20, mog21, mog22, mog23, mog24, mog25, mog26}. Among these solutions, rotating black holes are of particular astrophysical significance. The Kerr solution \cite{mog27} in GR provides the standard description of rotating black holes and plays a crucial role in understanding various high-energy astrophysical phenomena. Observational evidence indicates that the supermassive black holes located at the centers of galaxies, such as M87 and the Milky Way, possess significant angular momentum and are therefore expected to be rotating \cite{mog28}. Motivated by these considerations, a Kerr-like rotating black hole solution has also been constructed within the framework of MOG \cite{mog17}. This solution has subsequently been investigated by several authors in different physical settings to explore its geometrical structure and astrophysical implications \cite{mog30, mog31, mog32}.

In astrophysical environments, many observational signatures of black holes arise from the behavior of matter in their vicinity, particularly in the form of accretion discs. Accretion discs serve as the primary mechanism through which black holes gain mass \cite{mog33}. As matter spirals toward the event horizon of a black hole, viscous heating generates electromagnetic radiation across a broad range of wavelengths including x-rays, enabling the observation of black holes \cite{mog34,Bambi2012_PhysRevD.85.043002}. These accretion processes also power highly energetic phenomena such as active galactic nuclei and relativistic jets \cite{mog35}. Therefore, the extreme gravitational environment around accreting black holes provides a fundamental laboratory for testing GR and alternative theories of gravity. A substantial body of research has explored different black hole spacetimes subjected to external magnetic fields, examining aspects such as the dynamics of null and timelike particles, as well as the associated astrophysical implications \cite{Narzilloev20b, Narzilloev21b, Narzilloev22, Narzilloev22b, Narzilloev2023b, Narzilloev2023d, Narzilloev2023e, Narzilloev2024b}.

Keeping in view the astrophysical significance of the observed jet power and the radiative efficiency of accretion disks around black holes \cite{mog36}, we explore the properties of rotating black holes in the MOG framework. In particular we investigate the astrophysical implications of rotating MOG black holes by applying it to several well-known stellar-mass black hole candidates, namely, A0620-00, H1743-322, XTE J1550-564, GRS1124-683, GRO J1655-40, and GRS1915+105 \cite{mog37a,mog37b,mog37c,mog37d,mog37e,mog37f}. 

In the Kerr–MOG framework, the geometry of the spacetime is characterized by three fundamental parameters: the black hole mass $M$, the spin parameter $a$, and the modified gravity parameter $\alpha$, which quantifies deviations from the predictions of GR \cite{mog17}. We analyze how the presence of the MOG parameter alters the spacetime structure and influences the properties of particle motion in the vicinity of the black hole, with particular emphasis on the location of the ISCO. Since the ISCO plays a crucial role in the physics of accretion disk, any modification in its position directly affects observable quantities associated with accreting black holes. Within the framework of the Novikov–Thorne thin accretion disk model \cite{Narzilloev22a}, we show that the parameter $\alpha$ significantly modifies the radiative efficiency of accretion disks and introduces a degeneracy between $a$ and $\alpha$. This degeneracy implies that similar observational signatures may arise from different combinations of $a$ and $\alpha$. Using observational estimates of radiative efficiencies inferred from the continuum-fitting method (CFM), we constrain the allowed regions in the $(a, \alpha)$ parameter space for each source.

In addition, we investigate the power of relativistic jets using the Blandford–Znajek mechanism \cite{mog39}, and compare the theoretical predictions with the observed energetics of transient jets. In this analysis, we consider two representative values of the jet Lorentz factor, $\Gamma = 2$ and 
$\Gamma = 5$, to account for moderately and highly relativistic outflows. By combining the constraints obtained from both radiative efficiency and jet power, we identify the regions of the parameter space where the two observables can be simultaneously reproduced. Interestingly we show that, for several sources, significant overlap regions exist in the $(a, \alpha)$ parameter space, indicating that the Kerr–MOG model can consistently explain both the accretion and jet properties of these systems. However, for the highly spinning source GRS1915+105, compatibility is achieved only within a narrow range of Kerr–MOG parameters. 

We organize our work as follows. In the next section, we briefly review the Kerr–MOG black hole solution. In Sec. \ref{sec3}, we analyze the radiative efficiency and astrophysical relativistic jets for the Kerr–MOG black hole. Sec. \ref{sec4} is devoted to the study of a set of astrophysical black holes within the Kerr–MOG framework, including A0620–00, H1743–322, GRS 1124–683,  GRS 1915+105, XTE J1550–564,  and GRO J1655–40. The conclusions of our work are presented in the final section.
 
\section{Spacetime Structure of the Kerr–MOG Black Hole}
\label{sec2}

In the Boyer-Lindquist coordinates the Kerr-MOG black hole metric, reads

\begin{align}
    ds^2=&-\left(\frac{\Delta - a^2 \sin^2\theta}{\Sigma}\right)dt^2+\frac{\Sigma}{\Delta}dr^2\nonumber\\-\nonumber&2a\sin^2\theta\left(1-\frac{\Delta - a^2 \sin^2\theta}{\Sigma}\right)dtd\phi+\Sigma d\theta^2\\+\nonumber&\sin^2\theta\left[\Sigma+a^2\sin^2\theta\left(2-\frac{\Delta - a^2 \sin^2\theta}{\Sigma}\right)\right]d\phi^2\ ,\\ \label{metric} 
\end{align} 

with
\begin{align}
    \Delta&=r^2-2(1+\alpha)Mr+a^2+\alpha(1+\alpha)M^2,\\ \Sigma&=r^2+a^2\cos^2\theta\ . 
\end{align}

\begin{figure*}[ht!]
    \centering
    \includegraphics[width=0.5\linewidth]{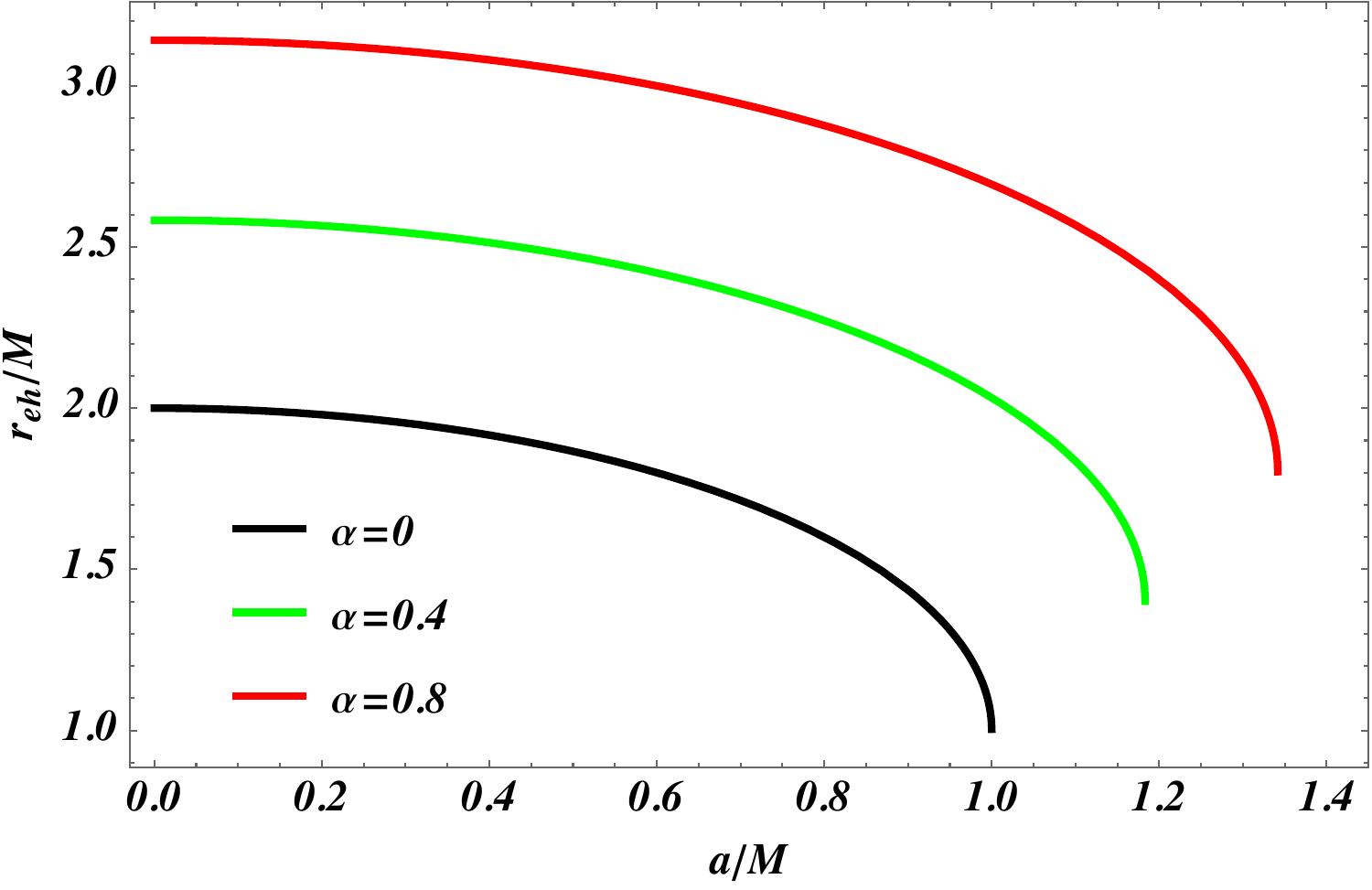}\includegraphics[width=0.5\linewidth]{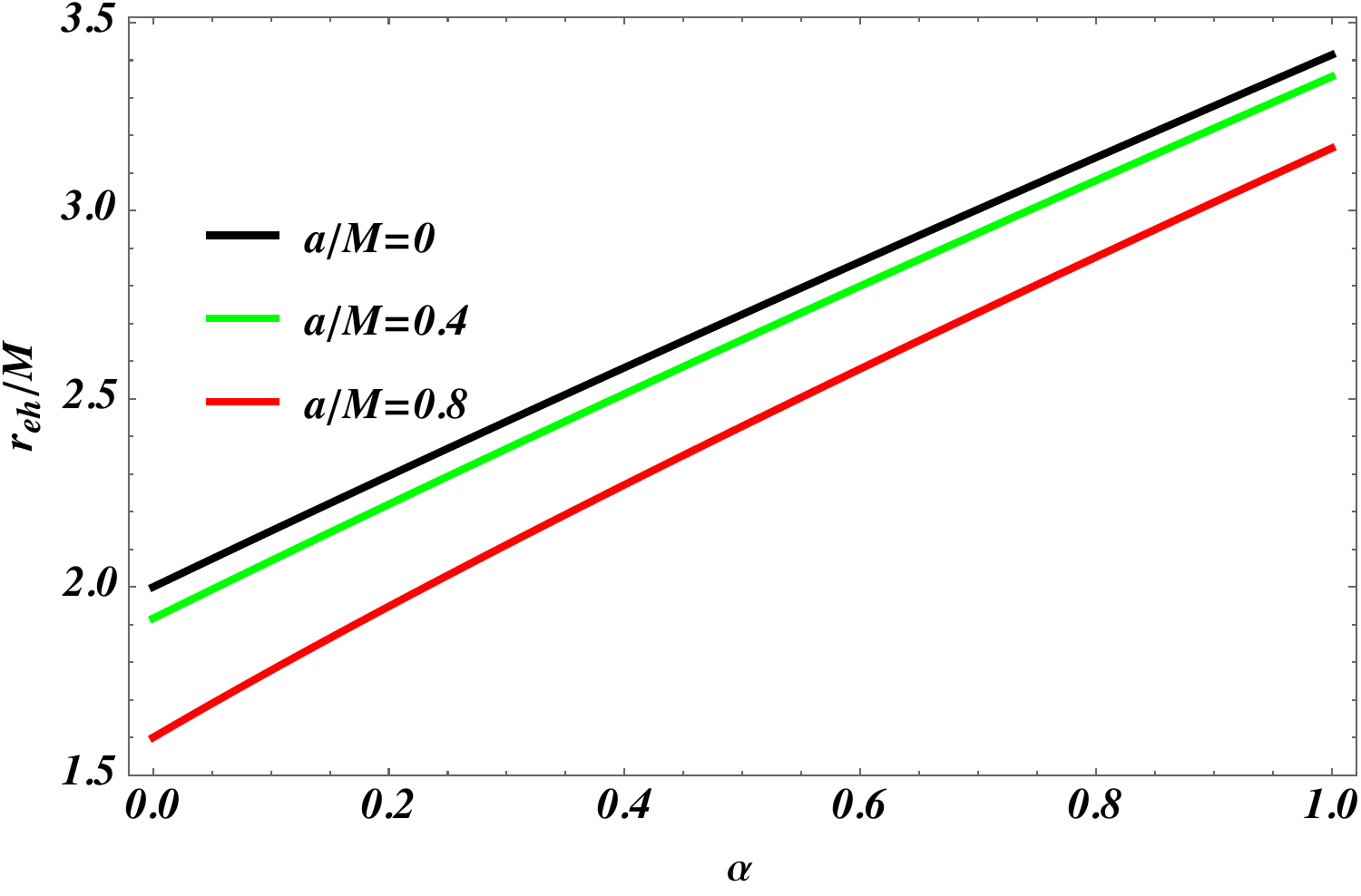}
    \caption{Variation of the horizon radii as functions of the spin parameter $a$ and the deformation parameter $\alpha$. The black solid line represents the extremal case characterized by coincident (degenerate) horizons.}
    \label{horizon_radiuses}
\end{figure*}

In what follows, we aim to investigate the properties of the spacetime metric (\ref{metric}) in order to demonstrate the degeneracy between its parameters and those characterizing the Kerr black hole. In particular, we analyze the horizon-like surfaces to assess how the MOG parameter $\alpha$ influences their structure. The horizons of a rotating Kerr--MOG black hole are determined by solving the following condition:
\begin{equation}
    g^{\sigma\rho}\partial_{\sigma}r\partial_{\rho}r=g^{rr}=\Delta=0\ .
    \label{horCase}
\end{equation}
This condition can be interpreted as defining a coordinate singularity of the spacetime metric (\ref{metric}). Numerical investigations indicate that, depending on the parameters $M$, $a$, and $\alpha$, Eq.~(\ref{horCase}) can yield up to two distinct positive real solutions, a degenerate (double) root, or no positive real solutions at all. These scenarios correspond to non-extremal black holes, extremal configurations, and the absence of black hole solutions, respectively, as shown in Fig.~\ref{horizon_radiuses}.\\ \indent When two positive real solutions exist, they represent the inner (Cauchy) horizon $r_{-}$ and the outer (event) horizon $r_{+}$, satisfying $r_{-} \leq r_{+}$, as depicted in Fig.~\ref{horizon_radiuses}. For the rotating case ($a \neq 0$), Eq.~\eqref{horCase} simplifies to the following expressions:

\begin{equation}
    r_{\pm}=(1+\alpha)M \pm \sqrt{(1+\alpha)M^2-a^2}.
    \label{horizons}
\end{equation}

Figure~\ref{horizon_radiuses} illustrates how the horizon radii $r_{\pm}$ vary with the spin parameter $a$ and the MOG parameter $\alpha$. For a fixed value of $a$, an increase in $\alpha$ leads to a reduction in the outer (event) horizon radius $r_{+}$, while the inner (Cauchy) horizon radius $r_{-}$ expands.\\ \indent Observers situated outside the event horizon and having zero angular momentum with respect to an observer at infinity are nevertheless compelled to rotate in the same direction as the black hole. This behavior arises from the frame-dragging effect induced by the rotating spacetime \cite{chandrasekhar1998mathematical}.

\begin{figure}
    \centering
    \includegraphics[width=0.95\linewidth]{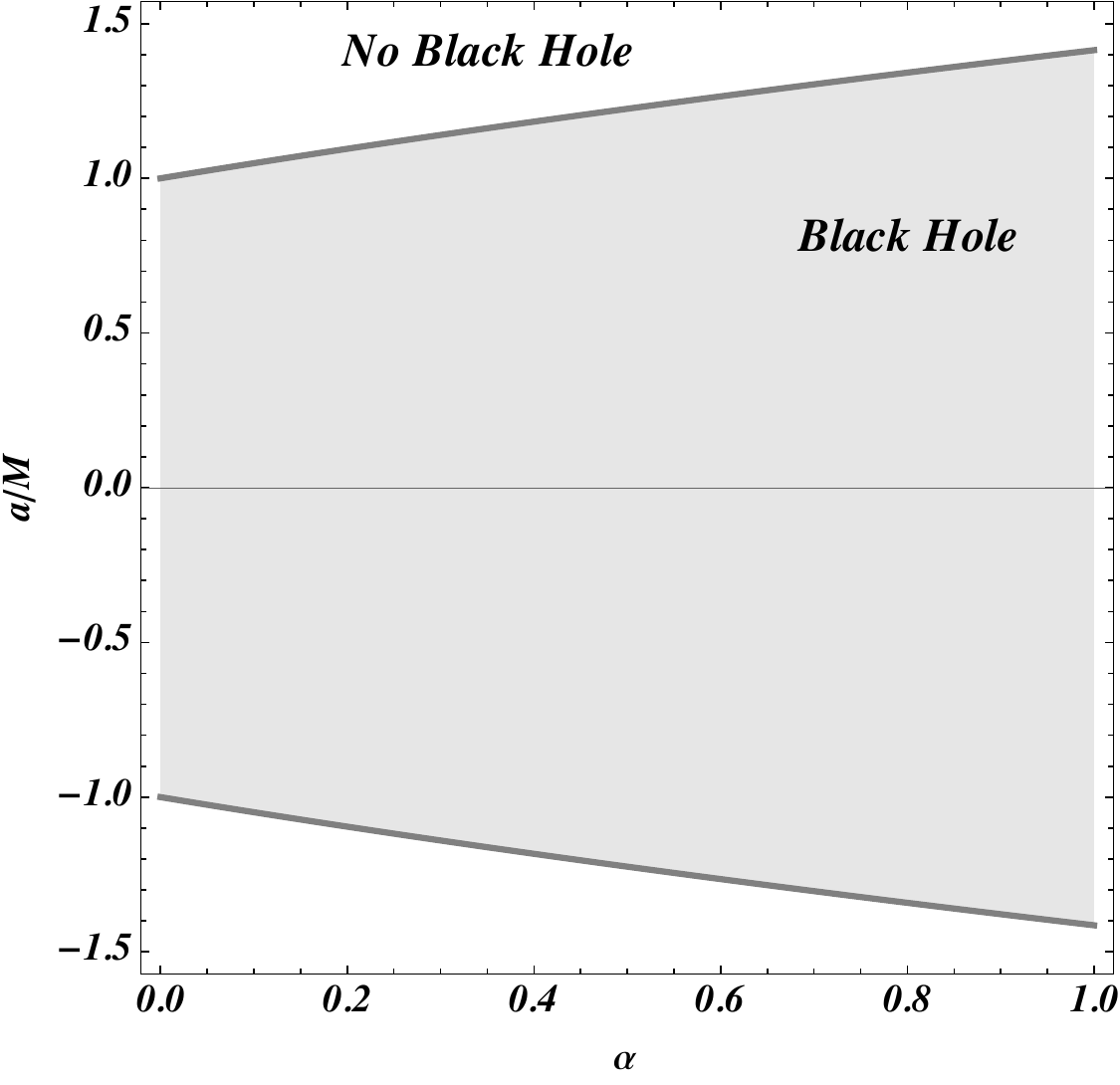}
    \caption{Allowed domain of the parameters ($a,\alpha$) corresponding to the existence of event horizons.}
    \label{regionHorizon}
\end{figure}

The conditions $\Delta = 0$ and $\partial_r \Delta = 0$ can be used to determine the boundary curves that distinguish black hole solutions from horizonless configurations. Figure~\ref{regionHorizon} presents the corresponding parameter space in terms of $(\alpha,a/M)$. Within the shaded (gray) region, the metric (\ref{metric}) possesses two distinct real roots, indicating the presence of a black hole. Outside this domain, no horizons are formed, and the spacetime does not describe a black hole. \\ \indent The solid black curves mark the set of parameter values for which the roots coincide, corresponding to the extremal case where the black hole has degenerate horizons. For $0 \leq \alpha \leq 1$, both positive and negative spin configurations are allowed within the black hole region. 
In the Kerr limit ($\alpha=0$), the admissible spin range is bounded by $|a|\leq M$, recovering the standard GR result.
In contrast, the Kerr--MOG spacetime admits higher absolute spin values, with the upper bound increasing to $|a|\leq \sqrt{1+\alpha}M$, reflecting the enlargement of the physically viable parameter space due to modified gravity effects.

\section{Theoretical aspect}
\label{sec3}
\subsection{Radiative efficiency of the system} 

In this study, we analyze the continuum emission from accretion disks surrounding black holes within the framework of the Novikov--Thorne thin-disk formalism, as previously applied in related works (see for example \cite{Narzilloev2023g,Saidov:prd}). This model offers a relativistic description of geometrically thin, optically thick disks orbiting compact objects. Nevertheless, it typically relies on simplified assumptions that may not fully capture subtle observational effects, including gravitational redshift and Doppler-induced frequency modifications. In particular, a detailed treatment of how such frequency shifts appear in realistic astrophysical environments, especially in the context of high-precision timing measurements, remains relatively underexplored. \\ \indent For the X-ray binary systems considered here, the Novikov--Thorne approximation is appropriate since the observed spectral states correspond to thin-disk configurations. The radial temperature profile follows the standard scaling relation $T(r) \propto r^{-3/4}$, consistent with observational studies. Under these conditions, the contribution from the corona is negligible, and the dominant radiation originates from the optically thick, thermally emitting disk. \\ \indent Within the Kerr spacetime, the model is characterized by four primary parameters: the black hole mass $M$, the dimensionless spin parameter $a_*$, the mass accretion rate $\dot{M}$, and the viscosity coefficient $\varsigma$. A key assumption is that the disk’s self-gravity can be neglected, so it does not modify the underlying spacetime geometry. Additionally, radial energy transport is ignored, and both energy and angular momentum are assumed to be radiated locally from the disk surface via photons with wavelengths much smaller than the characteristic gravitational scale. Magnetic field effects are also omitted. The inner boundary of the disk is taken to coincide with the ISCO, a condition that is essential for determining black hole spin through both continuum-fitting and iron line techniques. \\ \indent Our analysis extends earlier studies of non-standard black hole geometries by providing independent constraints based on accretion disk emission and jet-related processes. The continuum-fitting approach, grounded in the Novikov--Thorne model, has already demonstrated consistency with observational X-ray spectra in estimating black hole spins within GR. Consequently, this framework serves as a reliable and physically motivated tool for investigating possible deviations from GR, such as those introduced by modified gravity parameters. By incorporating these elements, the model facilitates a more realistic interpretation of observed spectra and connects theoretical thin-disk predictions with high-energy astrophysical observations. \\ \indent The disk material is assumed to be in local thermal equilibrium, resulting in emission that closely resembles a blackbody spectrum at each radius. For stellar-mass black holes, this radiation typically peaks in the soft X-ray band, whereas for supermassive black holes, the emission maximum is generally found in the ultraviolet region. \\ \indent The thermal radiation emitted by an accretion disk is strongly influenced by the position of its inner edge. If this inner boundary is identified with the ISCO, and independent estimates of the black hole’s mass, distance, and disk inclination are known, one can apply spectral fitting methods to infer the ISCO location. Notably, the ISCO radius is not a fixed quantity; instead, it is determined by the geometry of the surrounding spacetime. Its value can be obtained through an analysis of the effective potential governing particle dynamics near the black hole. \\ \indent It is essential to recognize that the ISCO is fundamentally linked to the specific form of the spacetime metric, and therefore varies across different gravitational models. Assuming a stationary and axially symmetric background described by a general metric $g_{\mu\nu}$, the ISCO can be derived by imposing the conditions required for circular motion and marginal stability of particle orbits \cite{Zhang_1997}. These conditions provide a systematic way to determine the critical radius at which stable circular motion ceases. Furthermore, enforcing the normalization condition for the particle four-velocity, $u_\mu u^\mu = -1$, leads directly to the corresponding equations of motion.%
\begin{equation}
    g_{rr} u_r^2 + g_{\theta\theta} u_\theta^2 = V_{\text{eff}} ,
\end{equation} 
where $V_{\text{eff}}$ is the effective potential that reads as \cite{bambi2017black}
\begin{equation}
    V_{\text{eff}} = \frac{E^2 g_{\phi\phi}+2 E L g_{t\phi}+L^2 g_{tt}}{g_{t\phi}^2-g_{tt}g_{\phi\phi}}-1\ .
\end{equation}
The specific energy and angular momentum of the test particle is depicted with $E$ and $L$ on the orbiting lines around black hole, respectively. In terms of the given spacetime metric they can be written as
\begin{equation}
    E= \frac{-g_{tt}- \Omega g_{t \phi}}{\sqrt{-g_{tt}-2 \Omega g_{t\phi} - \Omega^2 g_{\phi\phi}}},
\end{equation}
\begin{equation}
    L=\frac{g_{t\phi}+ \Omega g_{\phi\phi}}{\sqrt{-g_{tt}-2 \Omega g_{t\phi} - \Omega^2 g_{\phi\phi}}}\ ,
\end{equation}
where
\begin{equation}
    \Omega=\frac{d\phi}{dt}=\frac{-g_{t\phi,r}\pm\sqrt{(g_{t\phi,r})^2-(g_{\phi\phi,r})(g_{tt,r})}}{g_{\phi\phi,r}},
\end{equation}
denotes the angular velocity of the particle, as mentioned in \cite{bambi2017black} and $g_{\epsilon\rho,\sigma}=\partial_{\sigma}g_{\epsilon\rho}$.
By solving the following set of equations, the behavior of ISCO radius can be determined :
\begin{equation*}
    V_{\text{eff}}(r)=0, \qquad V_{\text{eff}}'(r)=0, \qquad V_{\text{eff}}''(r)=0.
\end{equation*}
\\ \indent It should be emphasized that the criteria used to identify the ISCO are directly governed by the specific components of the spacetime metric. As a result, observational estimates of the ISCO, particularly those obtained from the thermal continuum emission can be employed to place meaningful constraints on the parameters characterizing the gravitational field around the compact object. For instance, if the spacetime is modeled by the Kerr solution \cite{zhang2025black}, the ISCO location provides a direct means of inferring the black hole spin. This technique, commonly referred to as the CFM, has been extensively applied over the past two decades to determine the spins of stellar-mass black holes \cite{mcclintock2014black}.
\begin{figure*}[ht]
    \centering
\includegraphics[width=0.48\linewidth]{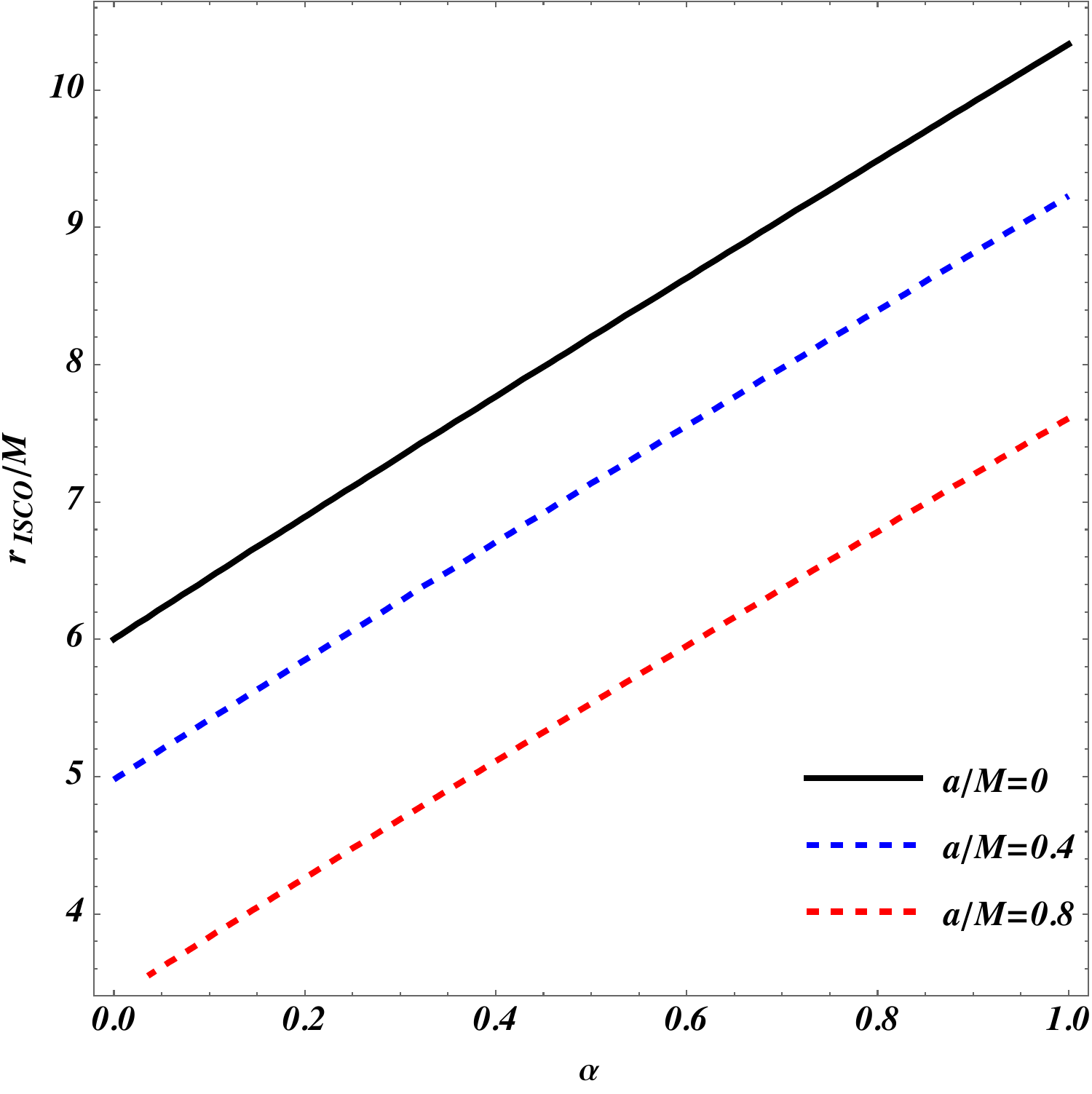}\includegraphics[width=0.48\linewidth]{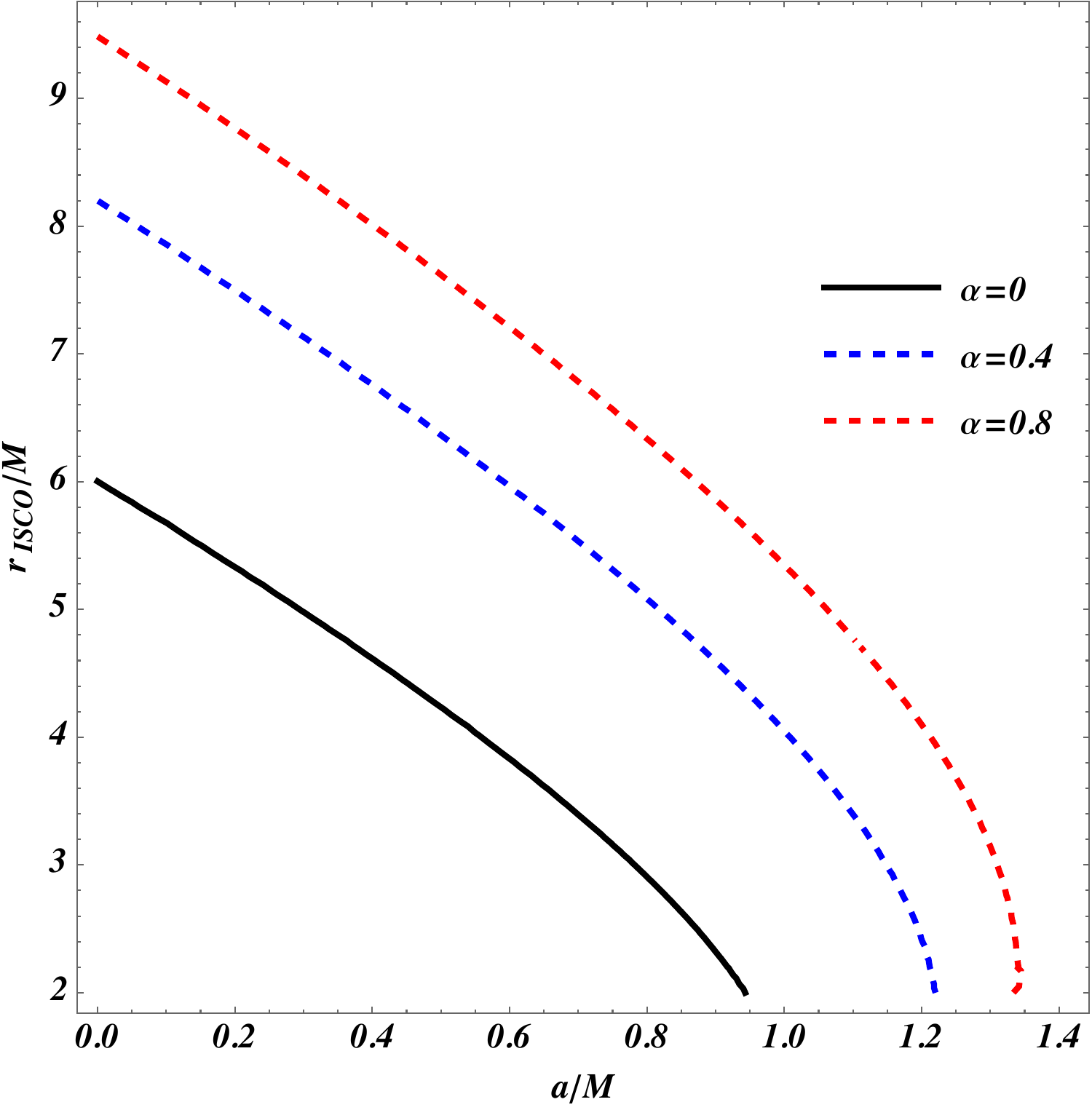}
    \caption{The ISCO radius of test particles as a function of the spin parameter $a$ and the MOG parameter $\alpha$ in the Kerr--MOG spacetime.
}
    \label{isco radiuses}
\end{figure*}

\begin{figure*}
    \centering
    \includegraphics[width=0.5\textwidth]{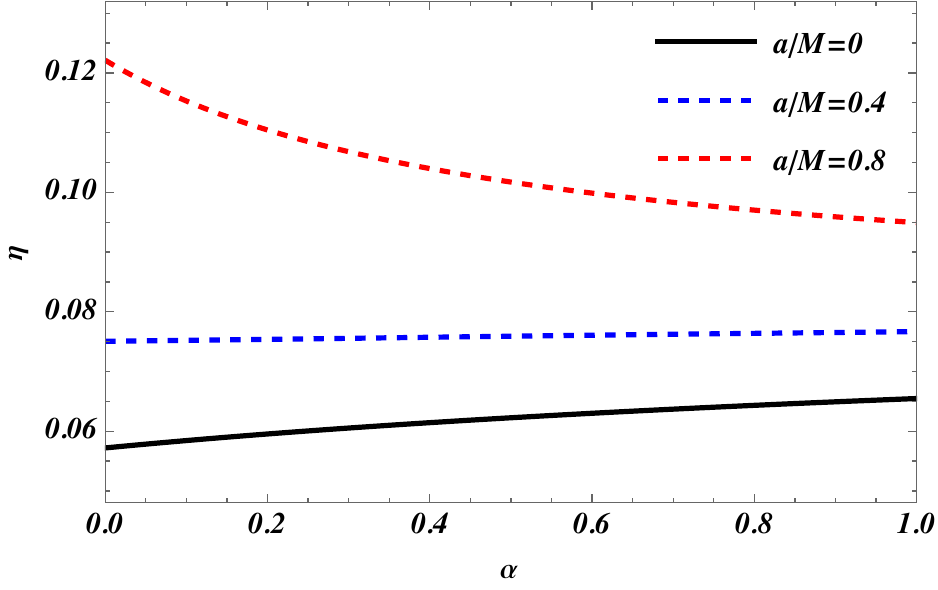}\includegraphics[width=0.5\textwidth]{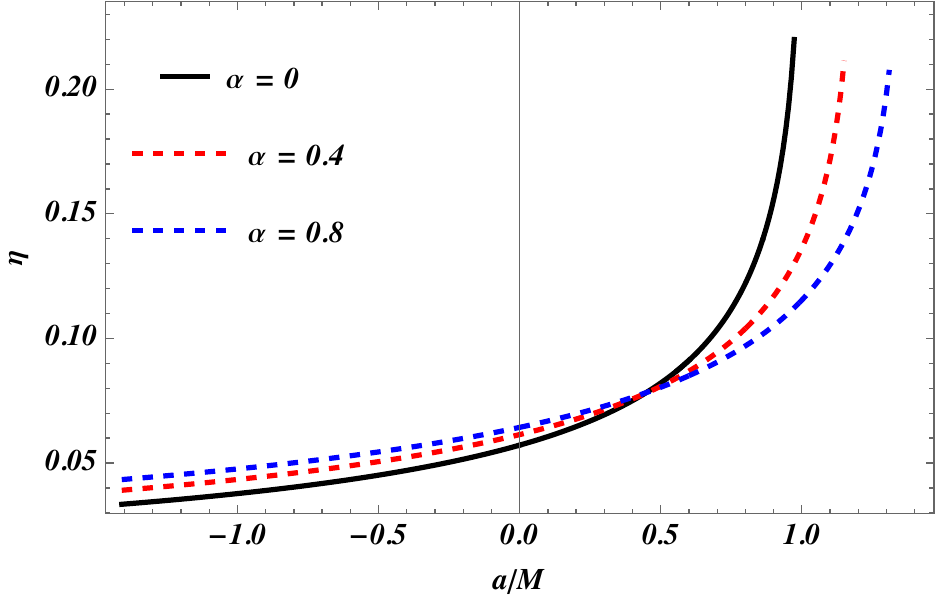}
    \caption{The radiative efficiency $\eta$ of a spinning Kerr--MOG black hole as a function of the spin parameter $a$ and the MOG parameter $\alpha$.
}
    \label{radEff}
\end{figure*}

Figure~\ref{isco radiuses} (left panel) illustrates that, for fixed values of spin, the ISCO radius increases in an approximately monotonic and nearly linear manner with the MOG parameter $\alpha$. In contrast, the right panel shows that, at a fixed $\alpha$, the ISCO radius decreases as the spin parameter $a$ grows, highlighting the familiar influence of rotation in reducing the radius of stable circular orbits in spinning black hole spacetimes.\\ \indent These two effects arising from the black hole spin $a$ and the MOG parameter $\alpha$ act in opposite directions, introducing degeneracies in their observational signatures. Such degeneracies can only be resolved by combining constraints from both accretion disk properties and jet-related observables.\\ \indent The radiative efficiency, denoted by $\eta$, of a Novikov--Thorne accretion disk is defined by the binding energy of a particle evaluated at the ISCO. This relation can be written as follows:
\begin{equation}\label{efficiency}
    \eta=1-E_{ISCO}.
\end{equation}
In this context, $E_{\rm ISCO}$ represents the specific energy of a particle evaluated at the ISCO. Accordingly, the radiative efficiency $\eta$ is fully determined by the properties of the background spacetime. In the case of the Kerr geometry, $\eta$ depends exclusively on the spin parameter $a$, whereas within the Kerr--MOG scenario it is influenced by both the spin $a$ and the modified gravity parameter $\alpha$.\\ \indent As a consequence, the Kerr--MOG framework leads to altered efficiency profiles, as shown in Fig.~\ref{radEff}. The left panel indicates that, for relatively small values of the spin parameter $a$, $\eta$ exhibits a modest increase with growing $\alpha$. In contrast, at higher spin values, the efficiency initially attains larger magnitudes but gradually declines as $\alpha$ increases, revealing a complex interplay between rotational effects and modifications to gravity.\\ \indent The right panel further illustrates that, at low spin, $\eta$ remains small, beginning from the lower bound $a = -\sqrt{1+\alpha}$, and increases progressively with increasing spin. In certain regions of parameter space, the efficiency shows a rapid, nearly divergent growth for larger values of the spin, emphasizing its strong sensitivity to rotation in the Kerr--MOG framework. This behavior implies that, in some intervals, $\eta$ is only weakly dependent on $a$, while in others a single spin value may correspond to a broad range of efficiencies.\\ \indent As depicted in Fig.~\ref{radEff}, within the context of the Shakura--Sunyaev disk model \cite{34,35}, two characteristic regimes emerge. The first corresponds to a weak dependence on spin, where $\eta$ remains approximately constant over the range $-\sqrt{1+\alpha} \lesssim a \lesssim 1.2$. The second regime occurs for $a \gtrsim 1.2$, where the efficiency increases rapidly, reaching values in the approximate interval $0.1 \lesssim \eta \lesssim 0.26$. This highlights the pronounced sensitivity of radiative efficiency to rapidly rotating black holes in the Kerr--MOG scenario.\\ \indent At the same time, the inclusion of the MOG parameter moderates the growth of $\eta$ and effectively imposes an upper bound on the allowed spin. Within the Novikov--Thorne framework, black holes sharing the same radiative efficiency are expected to produce nearly identical thermal spectra \cite{36}. Therefore, measurements of $\eta$ provide a valuable observational probe for testing deviations from the standard Kerr geometry.

\subsection{Astrophysical relativistic jets}
\begin{figure*}
    \centering
    \includegraphics[width=0.5\linewidth]{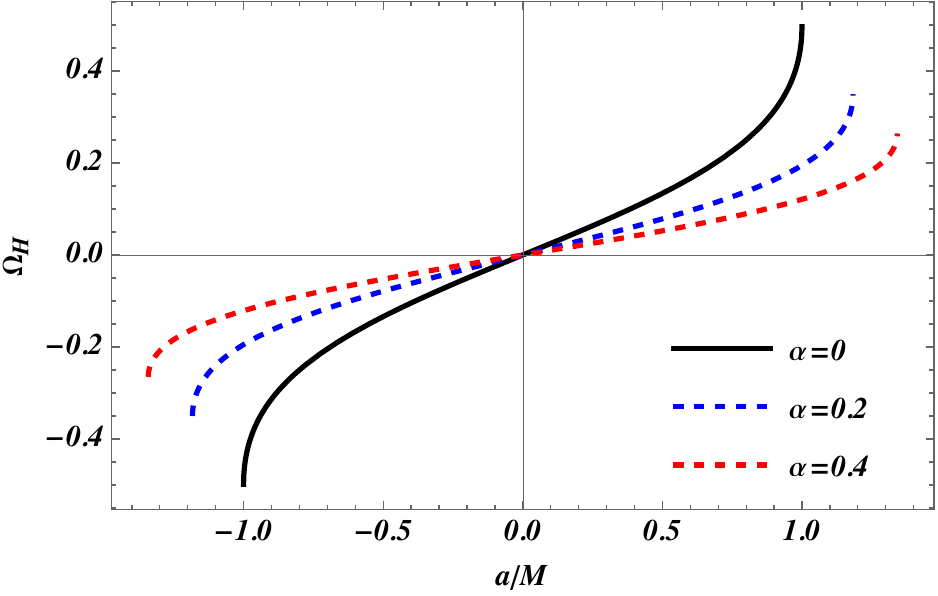}\includegraphics[width=0.5\linewidth]{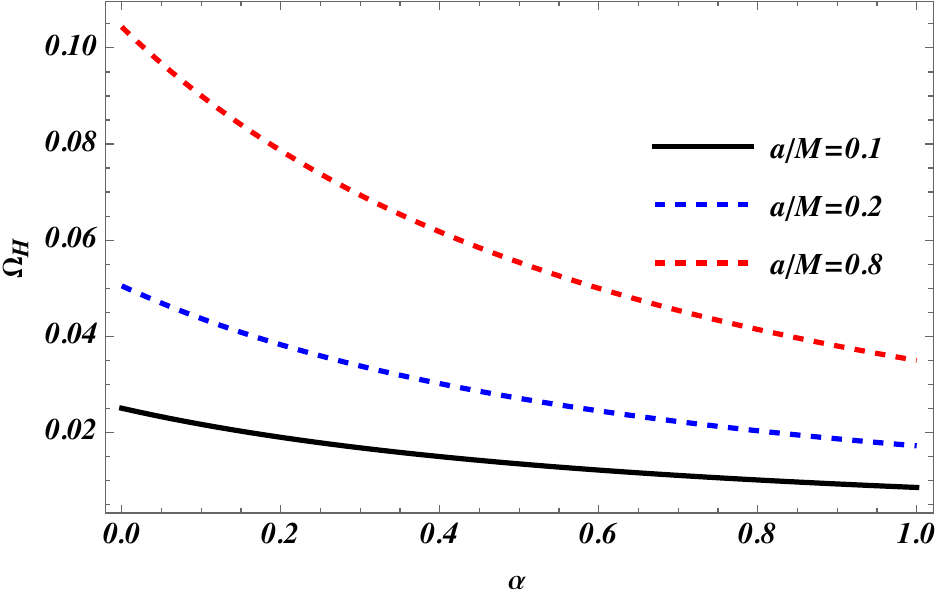}
    \caption{Variation of the horizon angular velocity $\Omega_H$ with the spin parameter $a$ for different values of the MOG parameter $\alpha$.
}
    \label{angVelHor}
\end{figure*}

Microquasars are known to exhibit two primary classes of jet outflows \cite{37}. The first category consists of persistent, mildly relativistic jets that are typically associated with the hard spectral state \cite{38} and can exist across a wide span of accretion luminosities. The second category includes transient, or ballistic jets, which are launched during state transitions when the luminosity nears the Eddington limit, particularly in the evolution from the hard state to the soft state. These transient jets are generally highly relativistic and are believed to originate in the vicinity of the event horizon \cite{39}.\\ \indent In the present work, our attention is directed toward transient jets, since their observational properties are especially sensitive to both the black hole spin and the MOG parameter $\alpha$, thereby providing valuable insight into the structure of the underlying spacetime. Despite extensive studies on the formation of relativistic jets \cite{40,41}, a fully consistent theoretical framework that accounts for all observed features is still lacking.\\ \indent To model jet production, we adopt the Blandford--Znajek mechanism \cite{42}, which describes the extraction of rotational energy from a spinning black hole. This process is applicable to any stationary, axisymmetric spacetime and explains how relativistic outflows can be driven by magnetic fields threading the event horizon and connected to a conducting accretion disk. Within this framework, the energy--momentum tensor is assumed to be dominated solely by electromagnetic contributions, while other forms of matter or fields are neglected.
\begin{equation}
    T_{\mu\nu}^{tot}\simeq T_{\mu\nu}^{EM} = F_{\mu\nu}F_\mu^\nu - \frac{1}{4}g_{\mu\nu}F_{\alpha\beta}F^{\alpha\beta}.
\end{equation}
By the conservation equation, it is simplified to 
\[
\nabla^\mu T_{\mu\nu}^{EM}=0,
\]
where, the electromagnetic field tensor is defined as $F_{\mu\nu} = A_{\nu,\mu} - A_{\mu,\nu}$, where $A_\mu$ denotes the four-potential. Within a force-free magnetospheric environment, one can directly obtain the following relation:
\begin{equation}
    \frac{A_{t,r}}{A_{\phi,r}}=\frac{A_{t,\theta}}{A_{\phi,\theta}}=-\omega(r,\theta).
    \label{13}
\end{equation}
In this expression, $\omega(r,\theta)$ represents the angular velocity associated with the electromagnetic field \cite{42}. By imposing condition (\ref{13}) on the electromagnetic four-potential, the corresponding field tensor $F_{\mu\nu}$ can then be written as
\[
F_{\mu\nu} = \sqrt{-g}
\begin{pmatrix}
0 & -\omega B^\theta & \omega B^r & 0 \\
\omega B^\theta & 0 & B^\phi & -B^\theta \\
-\omega B^r & -B^\phi & 0 & B^r \\
0 & B^\theta & -B^r & 0
\end{pmatrix}.
\]
Within this framework, the power associated with relativistic jets can be expressed as \cite{42}:
\begin{equation}
    P_{BZ} = 4 \pi \int_0^{\pi/2} \sqrt{-g}T_t^rd\theta,
\end{equation}
Here, $T_t^{\;r}$ represents the radial component of the Poynting flux. It is assumed that the jet originates in the vicinity of the event horizon. The radial Poynting flux can be written as

\[
T^r_t = 2 r_H M \sin^2\theta (B^r)^2 \omega \left[\Omega_H - \omega\right] \bigg|_{r = r_H}.
\]
The angular velocity $\Omega_H$ calculated on the event horizon $r_H$ and expressed as
\[
\Omega_H = -\frac{g_{t\phi}}{g_{\phi\phi}}\bigg|r_H .
\]
It is worth emphasizing that the original work by Blandford and Znajek focused on the slow-rotation regime, where the spin parameter $a$ is small, establishing that the jet power scales as $a^2$ \cite{42}. 
Subsequent studies by Tchekhovskoy et al. \cite{43} generalized this result to nearly the full range of Kerr black hole spins, confirming that, to leading order, the jet power is proportional to the square of the horizon angular velocity, $\Omega_H^2$, in accordance with the Blandford--Znajek mechanism

\begin{equation}
    P_{BZ}=\sigma \Phi_{tot}^2 \Omega_H^2.
    \label{jet}
\end{equation}
In this context, $\sigma$ takes the value $1/6\pi$ for a split-monopole magnetic field configuration and $\sigma = 0.044$ for a paraboloidal geometry \cite{42}. 
The results reported in \cite{43} were originally derived within the Kerr spacetime; here, we adopt their validity even when generalizing to a modified Kerr--MOG background. 
In Eq.~(\ref{jet}), $\Phi_{\rm tot}$ denotes the total magnetic flux threading the horizon and is given by

\begin{equation}
 \Phi_{tot}=2\pi\int_0^\pi\sqrt{-g}|B^r|d\theta.
\end{equation}
In Fig.~\ref{angVelHor}, the angular velocity at the event horizon $\Omega_H$ is shown as a function of the rotation parameter $a$ for various strengths of the MOG parameter $\alpha$, including the Kerr limit.


\section{Observational Constraints on Model Parameters}
\label{sec4}
Since the radiative efficiency, as defined in Eq.~\eqref{efficiency}, is fundamentally shaped by the spacetime geometry, it provides a sensitive probe of black hole parameters in modified gravity scenarios. 
Here, we examine a set of astrophysical black holes within the Kerr–MOG Rotating black hole solution framework, including A0620–00, H1743–322, GRS 1124–683,  GRS 1915+105, XTE J1550–564,  and GRO J1655–40 \cite{44}.

The reported spin values for these sources are derived using the CFM, which is valid when the black hole is in the thermal dominant (soft) state and the accretion luminosity lies roughly between 5\% and 30\% of the Eddington limit. 
Under these conditions, the accretion disk is accurately modeled by the Novikov--Thorne thin-disk framework, with the inner edge located at the ISCO and negligible contributions from outflows or jets.

In this regime, the disk's accretion luminosity is given by
\begin{equation}
    L_{\rm acc} = \eta_r \, \dot{M} c^2,
\end{equation}
where the radiative efficiency is defined by Eq.~\eqref{efficiency}. 
Within this framework, both $L_{\rm acc}$ and $\eta_r$ are determined solely by the black hole spin parameter.

In the presence of outflows (excluding relativistic jets), the total luminosity can be expressed as
\begin{equation}
    L_{\rm acc} = (\eta + \eta_k)\,\dot{M}c^2,
\end{equation}
where $\eta_k$ denotes the fraction of energy transported away in the form of kinetic outflow power. Under these conditions, the effective radiative efficiency is diminished, such that $\eta < 1 - E_{\rm ISCO}$.\\ \indent It is important to note that black hole spin can be estimated using approaches other than the CFM. One such technique is the outflow-based method, which has been applied to both stellar-mass and supermassive black holes \cite{45,46}. However, spin values inferred from this approach often show inconsistencies when compared with those obtained via CFM. For instance, in the case of A0620--00, the CFM yields a relatively low spin value of $a = 0.12 \pm 0.19$ \cite{47}, while the outflow method suggests a significantly larger spin \cite{45}.\\ \indent In this work, we rely solely on spin measurements derived from the CFM, assuming their reliability. Table~\ref{table1} lists the adopted spin parameters together with the corresponding Novikov--Thorne radiative efficiencies $\eta$, all evaluated within the Kerr spacetime. The parameters reported in Table~\ref{table1} are compiled from previously published continuum-fitting studies of black hole X-ray binaries, largely based on data from the Rossi X-ray Timing Explorer (RXTE) \cite{47,48,49,50,51,52}.\\ \indent These analyses provide well-constrained estimates of black hole mass, spin, and disk inclination by fitting thermal X-ray spectra within the Novikov--Thorne thin-disk framework. The dominant sources of systematic uncertainty include errors in the measurements of mass, distance, and inclination, as well as possible deviations from the thin-disk approximation and instrumental calibration limitations. In the present study, these observational inputs are used to test and constrain the theoretical predictions of the Kerr--MOG model.
\\ \indent In the present work, we follow the formalism developed in Refs.~\cite{Narzilloev2023g,Saidov:prd} to evaluate the jet power for the six astrophysical sources considered. 
Within this framework, a bipolar radio jet is modeled as a symmetric pair of plasmoids, which emit radiation isotropically and exhibit a narrow optical structure \cite{Saidov}. 
These plasmoids propagate away from the central engine with a relativistic bulk velocity $\beta$. 
The ratio between the observed and intrinsic flux densities of each jet component is given by
\begin{equation}
    \frac{S_\nu}{S_{\nu,0}} = \delta^{3-\varphi}.
\end{equation}
Here, $\delta$ denotes the Doppler boosting factor, while $\varphi$ represents the radio spectral index. 
For the approaching jet, the Doppler factor can be written in terms of the bulk velocity $\beta$, the Lorentz factor $\Gamma$, and the jet inclination angle $i$ as
\[
\delta = \left[\Gamma(1-\beta \cos i)\right]^{-1}.
\]

\begin{widetext}

\begin{table}[h!]
\centering
\caption{Properties of the binary black hole systems analyzed in this study. The radiative efficiency $\eta$ is computed from the measured spin using Eq.~(10) within the Kerr spacetime framework.
}
\begin{tabular}{lcccccc}
\toprule
\textbf{BH source} & $M\,(M_\odot)$ &\hspace{0.5cm} $D$ (kpc) &\hspace{1cm} $i$ ($^\circ$) &\hspace{0.6cm} $a$ & \hspace{1cm}$\eta$ \\
\midrule
A0620-00 &\hspace{0.4cm} $6.61 \pm 0.25$ &\hspace{0.4cm} $1.06 \pm 0.12$ &\hspace{1cm} $51.0 \pm 0.9$ &\hspace{1cm} $0.12 \pm 0.19$~\cite{47} & \hspace{1cm}$0.061^{+0.010}_{-0.008}$ \\
H1743-322 & $8.0$ &\hspace{0.2cm} $8.5 \pm 0.8$ & \hspace{1cm}$75.0 \pm 3.0$ &\hspace{0.8cm} $0.20 \pm 0.3$~\cite{48} & \hspace{1cm}$0.065^{+0.017}_{-0.015}$ \\
XTE J1550-564 &\hspace{0.4cm} $9.10\pm0.61$ & \hspace{0.6cm}$4.38 \pm 0.58$ &\hspace{1cm} $74.7 \pm 3.8$ &\hspace{1cm} $0.34 \pm 0.24$~\cite{49} &\hspace{0.9cm} $0.270^{+0.120}_{-0.100}$ \\
GRS 1124-683 & $11.0^{+2.1}_{-1.4}$ &\hspace{0.2cm} $4.95^{+0.69}_{-0.65}$ &\hspace{1cm} $43.2 \pm 2.1$ &\hspace{0.7cm} $0.63^{+0.16}_{-0.19}$~\cite{50} &\hspace{1cm} $0.095^{+0.025}_{-0.030}$ \\
GRO J1655-40 &\hspace{0.45cm} $6.30 \pm 0.27$ &\hspace{0.2cm} $3.2 \pm 0.5$ &\hspace{1cm} $70.2 \pm 1.9$ &\hspace{1cm} $0.70 \pm 0.10$~\cite{51} &\hspace{1cm} $0.104^{+0.018}_{-0.021}$ \\
GRS 1915+105 &\hspace{0cm} $12.4^{+2.0}_{-1.8}$ & \hspace{0.1cm}$8.6^{+2.0}_{-1.6}$ &\hspace{1cm} $60.0 \pm 5.0$ &\hspace{1cm} $a_* > 0.98$~\cite{52} &\hspace{1cm} $\eta = 0.234$ \\
\bottomrule
\end{tabular}
\label{table1}
\end{table}
\end{widetext}

%%%%
\begin{table}[h]
    \centering
    \caption{Proxy jet power for the selected sources, expressed in units of kpc$^2$ GHz Jy $M_{\odot}^{-1}$}
    \begin{tabular}{lccccc}
        \toprule
        BH source & $(S_{\nu,0})_{max}^{5GHz} (Jy)$ & $P_\text{jet} \bigg|_{\Gamma=2}$ & $P_\text{jet}\bigg|_{\Gamma=5}$ \\
        \midrule
        A0620-00 & 0.203 & 0.13 & 1.6 \\
        H1743-322 & 0.0346 & 7.0 & 140 \\
        XTEJ1550-564 & 0.265 & 11 & 180 \\
        GRSJ124-683 & 0.45 & 3.9 & 380 \\
        GROJ1655-40 & 2.42 & 70 & 1600 \\
        GRS1915+105 & 0.912 & 42 & 660 \\
        \bottomrule
    \end{tabular}
    \label{table2}
\end{table}

For the jet approaching the observer, Doppler boosting enhances the observed intensity at low inclination angles, while at larger inclinations the emission becomes suppressed. 
In microquasars with mildly relativistic jets, the Doppler factor typically drops below unity for inclinations of about $35^\circ$--$55^\circ$. 
We further assume that the total transient jet energy is proportional to the peak radio flux density at 5~GHz (see Table~\ref{table2}).

In natural units, the jet luminosity can be expressed as \cite{PhysRevD.10.1680,RPenrosePhysRevLett14571965}
\begin{equation}
    P_{\rm jet} = \left(\frac{\nu}{5~{\rm GHz}}\right)\left(\frac{S_{\nu,0}^{\rm tot}}{\rm Jy}\right)\left(\frac{D}{\rm kpc}\right)^2\left(\frac{M}{M_{\odot}}\right)^{-1} \, .
\end{equation}
Here, $S_{\nu,0}^{\rm tot}$ represents the combined flux density of the approaching and receding jet components \cite{39}. 
The jet Lorentz factor $\Gamma$ is typically expected to fall within the range $2 \leq \Gamma \leq 5$. 
The Doppler-corrected jet power values corresponding to $\Gamma = 2$ and $\Gamma = 5$ for each source are presented in Table~\ref{table2} \cite{44,53}.

The results presented in Table~\ref{table2} may be directly confronted with theoretical expectations, which depend on the underlying spacetime geometry. The jet power can be parametrized as
\begin{equation}
    \log P = \log K + 2 \log \Omega_H \, .
\end{equation}
Within this formulation, the coefficient $K = \sigma \Phi_{\rm tot}^2$ is obtained by fitting the observationally inferred jet powers to the corresponding values of the horizon angular velocity $\Omega_H$, following the procedure outlined in Ref.~\cite{53}. The influence of the spacetime metric enters explicitly through the dependence of the jet power on $\Omega_H^2$.
Observational constraints on jet power are provided by Ref.~\cite{53}, which gives best-fit normalizations of 
$\log K = 2.94 \pm 0.22$ and $\log K = 4.19 \pm 0.22$ (90\% confidence-level (CL)) 
for jet Lorentz factors of $\Gamma = 2$ and $\Gamma = 5$, respectively. While $K$ varies between sources, theoretical models suggest a correlation between magnetic field strength and accretion rate $\dot{M}$ \cite{Narzilloev22a,54}.
For ballistic jets, which are typically produced during transitions from the hard to the soft spectral state, the Eddington-normalized accretion rate $\dot{M}$ is expected to be nearly uniform across different sources. In addition, the black hole masses in these systems are of the same order, generally around $\sim 10\,M_\odot$. Under these conditions, the parameter $K$ can be treated as approximately constant for the six black hole candidates analyzed in this work. If one further assumes that $K$ does not depend on the underlying spacetime geometry, its fitted values can be used to place constraints on both the spin parameter $a$ and the MOG parameter $\alpha$ in the Kerr--MOG framework. These constraints are derived by comparing theoretical predictions with the observed jet power measurements listed in Table~\ref{table2}.

\subsection{Candidate A0620-00}
\begin{figure*}[ht]
    \centering
    \includegraphics[width=0.52\linewidth]{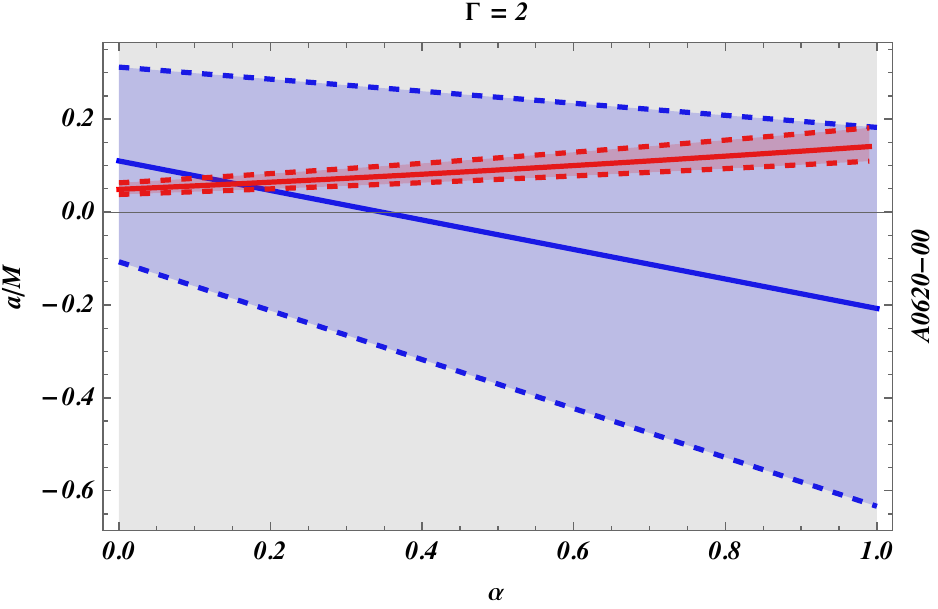}\includegraphics[width=0.52\linewidth]{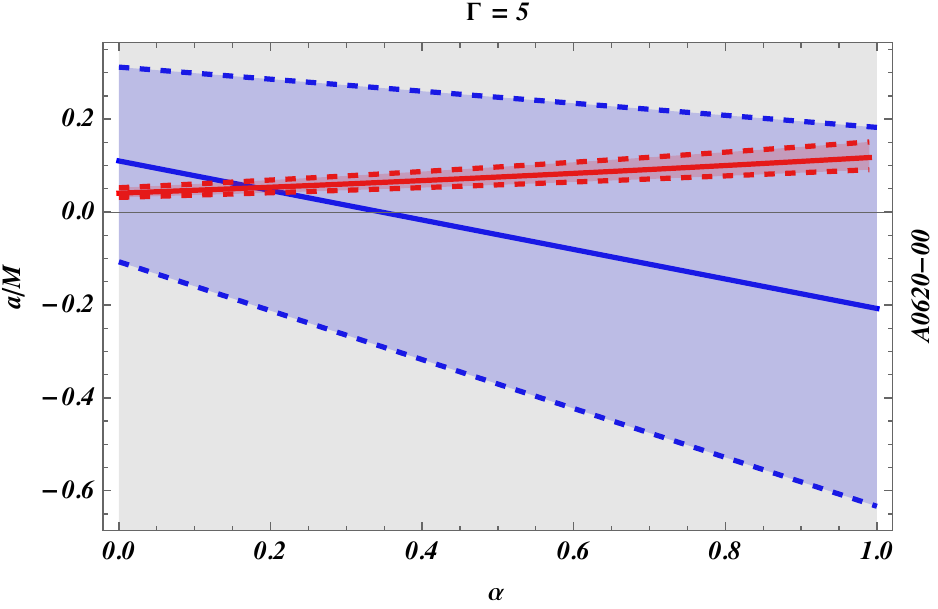}
    \caption{Constraints on the spin parameter $a$ and MOG parameter $\alpha$ for A0620--00. Blue curves and shaded regions correspond to the radiative efficiency $\eta$ and its uncertainties (Table~\ref{table1}), while red curves indicate jet power constraints with a 0.3 dex uncertainty. Left and right panels assume Lorentz factors $\Gamma = 2$ and $\Gamma = 5$, respectively. The overlap regions satisfy both thermal and jet constraints. Gray areas denote parameter regions admitting black hole solutions.}

    \label{s1}
\end{figure*}

The CFM applied within the Kerr metric framework, 
yields a spin estimate of $a = 0.12 \pm 0.19$ (68\% CL) for the chosen source \cite{47}. 
This corresponds to a radiative efficiency of $\eta = 0.061_{-0.007}^{+0.009}$   
(see Table~\ref{table1}). As noted in \cite{36}, the CFM provides a first-order 
estimate of the Novikov-Thorne disk efficiency, which can then be used to 
constrain possible deviations from Kerr spacetime. 
Here, we apply this method to constrain the parameters of the Kerr-MOG spacetime, 
specifically the spin $a$ and the MOG parameter $\alpha$, for rotating black holes. Figure \ref{s1} shows (in blue regions) the bounds on the spin parameter $a$ and parameter $\alpha$, based on the condition that the Novikov-Thorne radiative efficiency is  $\eta = 0.061_{-0.007}^{+0.009}$. For $ \alpha = 0 $, the spin parameter is  $a = 0.11 \pm 0.2$ , though greater spin values take when  $\alpha > 0$ . Due to a degeneracy in the radiative efficiency across the $(a, \alpha)$ parameter space, simultaneous constraints on both variables are ambiguous without further observational input. The left panel of Fig.~\ref{s1} corresponds to $\Gamma = 2$, and the right panel to $\Gamma = 5$.
In both, the solid red line marks the central value of $P_{\mathrm{jet}}$, 
while the dashed red lines show the ranges of spacetime parameters 
that produce jet powers within a $0.3$~dex margin around this central value 
(see Table~\ref{table2}). The shaded red areas show the range of spacetime parameters that fit within the error bars of the observations.
In both the left plot (for $\Gamma = 2$) and the right plot (for $\Gamma = 5$), 
the jet power of the source is well explained by the Kerr-MOG model. For $\Gamma = 2$ (left), the best-fit jet power matches what the standard Kerr spacetime predicts 
with spin $a \approx 0.05$. The Kerr-MOG model also fits the data well over a range of parameters. For $\Gamma = 5$ (right), we see a small shift compared to the $\Gamma = 2$ case, but the difference is minor and the Kerr-MOG model still agrees with the observations. It is apparent that increasing the MOG parameter $\alpha$ induces a slight enhancement in the black hole spin parameter $a$ within the Kerr–MOG spacetime framework. As illustrated in Fig.~\ref{s1}, the jet power trends corresponding to Lorentz factors $\Gamma = 2$ and $\Gamma = 5$ exhibit nearly identical behavior. The overlap between the blue and red shaded regions represents the range of spacetime parameters capable of simultaneously reproducing both observational constraints within their respective uncertainties under the Kerr–MOG model. Accordingly, the figure indicates that both the radiative efficiency and jet power of the source A0620--00 are consistently accounted for within the Kerr–MOG rotating black hole spacetime. The regions of the blue and red shaded contours constrains the spacetime parameters to approximately $0.04 < a < 0.06$ and $0 < \alpha < 1$. Moreover, the optimal parameter values that simultaneously reproduce the central observational estimates correspond to $a \simeq 0.04$ for the spin and $\alpha \simeq 0.16$ for the MOG coupling parameter. These values are located near the intersection of the best-fit radiative efficiency and jet power points.

\subsection{Candidate H1743-322}
\begin{figure*}[ht]
    \centering
    \includegraphics[width=0.5\linewidth]{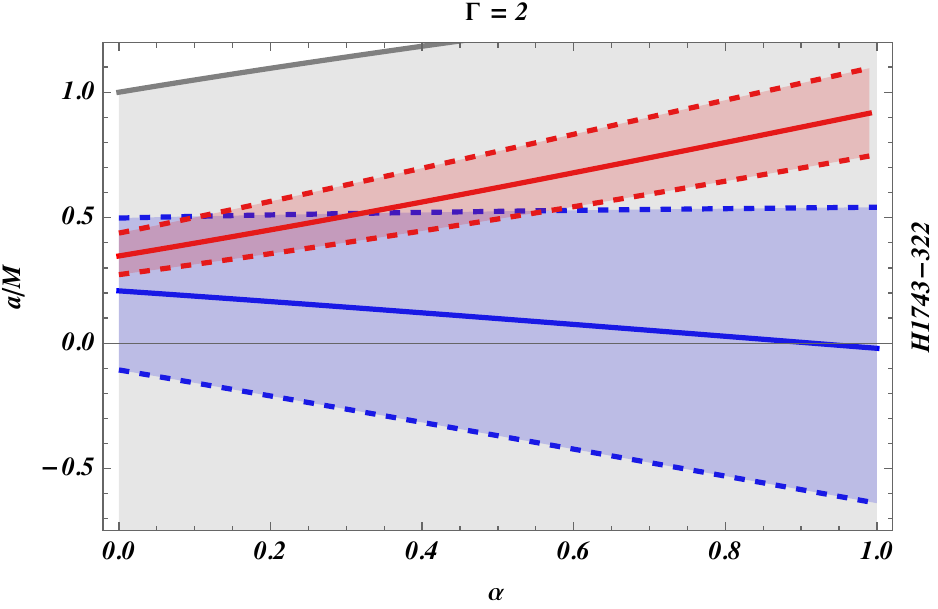}\includegraphics[width=0.5\linewidth]{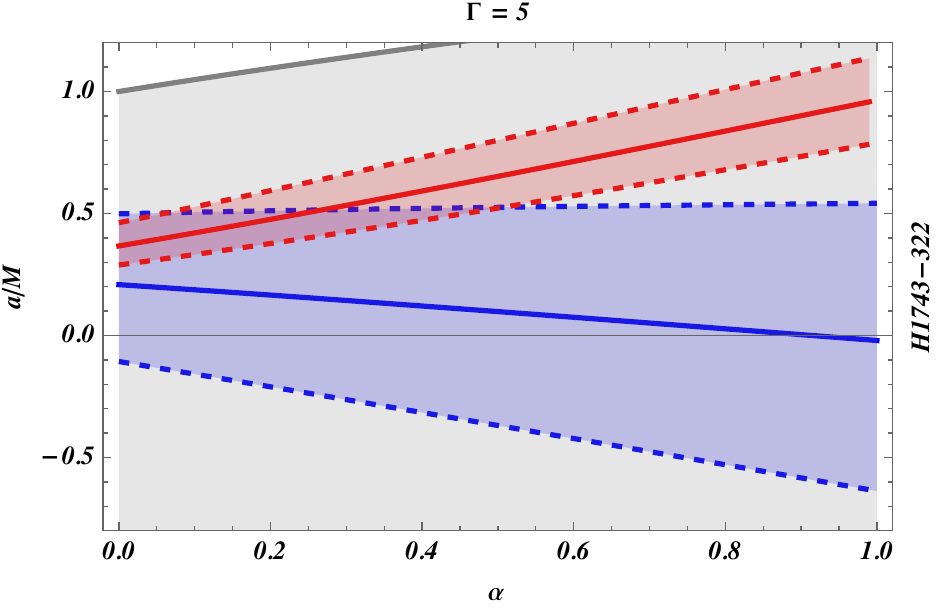}

\caption{For H1743-322, constraints on the spin parameter $a$ and MOG parameter $\alpha$ for the blue shaded region indicates values consistent with the observed thermal spectrum. As in the previous figure, red curves represent jet power constraints, with left and right panels corresponding to Lorentz factors $\Gamma = 2$ and $\Gamma = 5$, respectively.}\label{s2}
\end{figure*}

The spin of the black hole in H1743-322 was determined by the CFM, yielding a value of \( 0.2 \pm 0.3 \) at a 68\% CL and \(-0.3 < a < 0.7 \) at a 90\% CL. Consequently, the radiative efficiency of the source is \( \eta = 0.065_{-0.011}^{+0.017} \) at a 68\% CL. The blue shaded regions in Fig.~\ref{s2} represent the allowed Kerr–MOG spacetime parameter space consistent with the measured radiative efficiency of H1743--322. In the Kerr limit, corresponding to $\alpha = 0$, the central estimate of the spin parameter is found to be $a \approx 0.208$. As the MOG parameter increases to $\alpha = 1$, the central radiative efficiency contour shifts toward lower spin values, reaching approximately $a \approx -0.021$. The dashed blue curves indicate the observational boundaries constraining the admissible parameter region. The red shaded regions in Fig. \ref{s2} illustrate the parameter constraints derived from the jet power measurements of H1743--322. As the MOG parameter increases from $\alpha = 0$ to $\alpha = 1$, the spin interval required to reproduce the observed jet power within the associated uncertainties undergoes a substantial shift for the case $\Gamma = 2$ (left panel), evolving from $a \approx 0.35 \pm 0.09$ to $a \approx 0.91 \pm 0.2$. In contrast, the corresponding variation is comparatively weak for $\Gamma = 5$, as shown in the right panel. The central estimates of the radiative efficiency and jet power do not coincide within the black hole parameter space. However, the overlap between the blue and red shaded regions persists within the observational uncertainties up to approximately $\alpha \approx 0.6$, indicating that consistent spacetime parameter values can still be accommodated within the error margins. This indicates that the observational data for this source are only partially consistent with the theoretical predictions of our model. The figure highlights that H1743-322, described by the spacetime geometry of the spinning black hole in Kerr-MOG solution, has a good chance of simultaneously explaining both observations involving radiative efficiency and jet power of the source.

\subsection{Candidate XTE J1550-564}
\begin{figure*}[ht]
    \centering
    \includegraphics[width=0.5\linewidth]{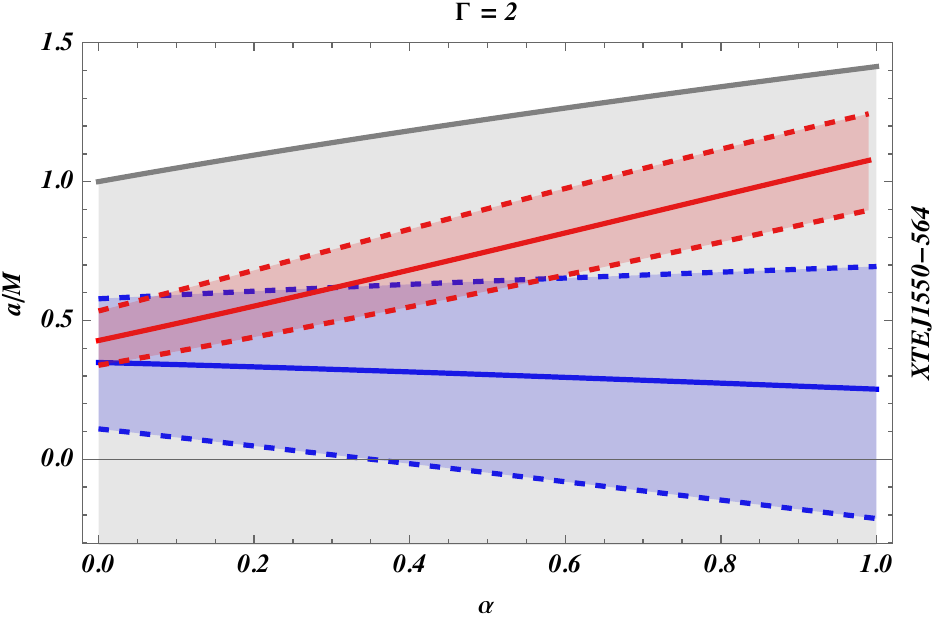}\includegraphics[width=0.5\linewidth]{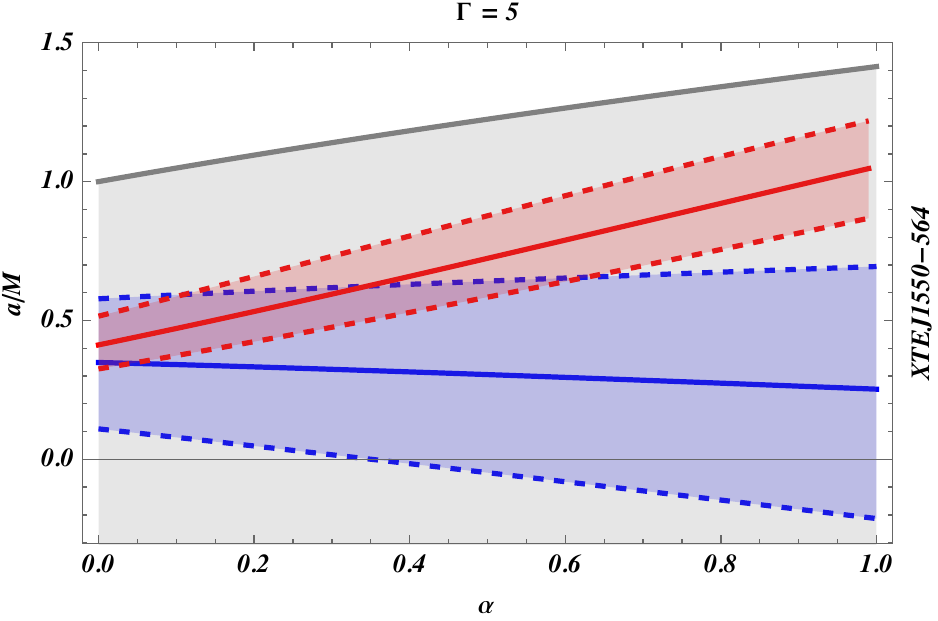}
    \caption{XTE J1550-564. The behavior observed in these diagrams is consistent with that of the previous case; a precise explanation is provided in the relevant discussion above.}
    \label{s3}
\end{figure*}

The spin estimate reported in \cite{49} yields  $a \simeq 0.34 \pm 0.15$ at the 68\% CL. In Fig.~\ref{s3}, we present the allowed parameter space in the  $a(\alpha)$ plane obtained by modeling the black hole spacetime within the Kerr--MOG framework. For the Kerr limit $\alpha = 0$, the radiative efficiency constraint corresponds to a central spin value of $a \simeq 0.35$, which gradually decreases to $a \simeq 0.239$ as the MOG parameter increases to $\alpha = 1$. The regions compatible with the observed jet power (shown in red) exhibit a quite similar behavior for both $\Gamma = 2$ and $\Gamma = 5$, remaining nearly unchanged acrosses the two cases. In contrast, the central spin values required to reproduce the jet power increase with the MOG coupling parameter, ranging from $a \simeq 0.428$ at $\alpha = 0$ up to $a \simeq 1.045$ at $\alpha = 1$. Consequently, the source XTEJ1550--564, previously interpreted as a Kerr black hole with spin $a \simeq 0.428 \pm 0.12$, can likewise be consistently described within the Kerr--MOG spacetime geometry. The optimal parameter values, which account for the observed $\eta$ and $P_{\text{jet}}$, correspond to the intersection region where the blue and red solid curves overlap, occurring at approximately $a\approx 0.67$ and $\alpha\approx0.646$

\subsection{Candidate GRS 1124-683}
\begin{figure*}[ht]
    \centering
    \includegraphics[width=0.5\linewidth]{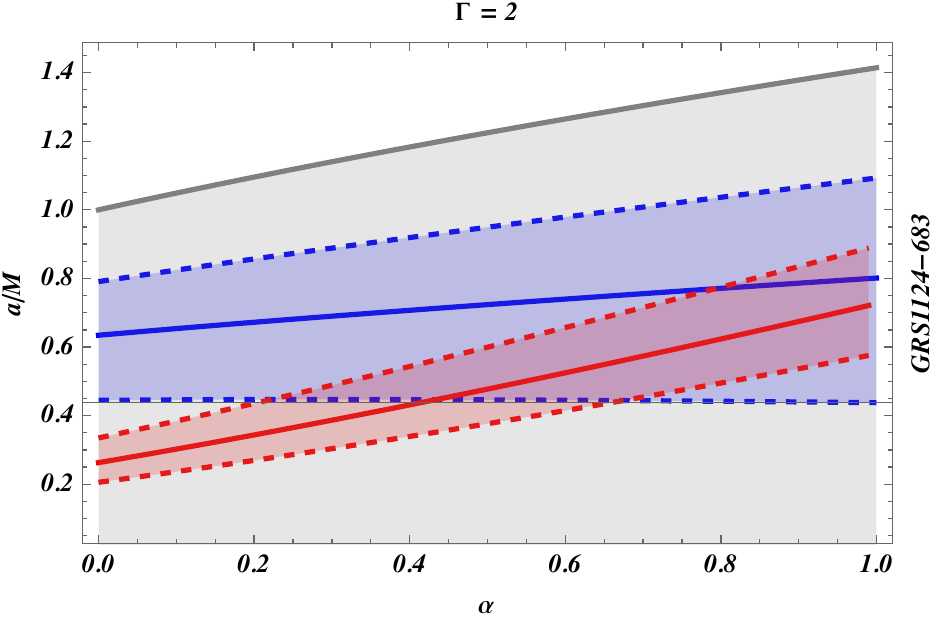}\includegraphics[width=0.5\linewidth]{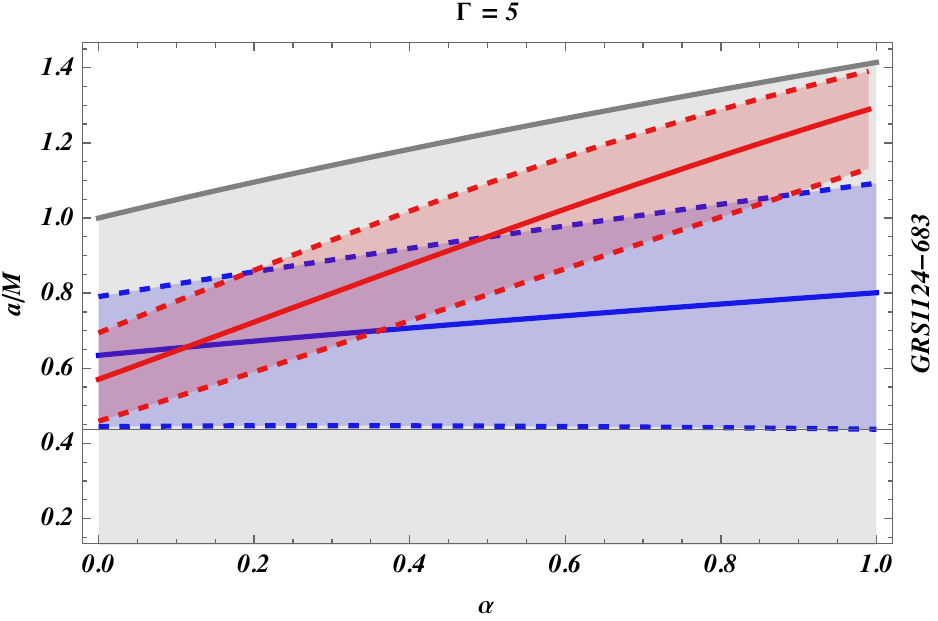}
    \caption{GRS 1124-683. The left panel demonstrates that, for a Lorentz factor $\Gamma = 2$, the Kerr--MOG spacetime cannot simultaneously reproduce the observed jet power and radiative efficiency of the source. In contrast, the right panel shows an overlap of the allowed regions for $\Gamma = 5$, indicating that the Kerr--MOG spacetime can reproduce the observations when the observational uncertainties are taken into account. Further discussion is provided in the main text.}
    \label{s4}
\end{figure*}

Using the CFM under the assumption of a Kerr background, the spin parameter of the black hole is measured to be $0.63^{+0.16}_{-0.19}$ at the 68\% CL \cite{50}. Based on this estimate, we derive the corresponding constraints on the spacetime parameters $a$ and $\alpha$, which are shown by the blue shaded regions in Fig.~\ref{s4} associated with the radiative efficiency. 
The central (solid) blue curve extends up to approximately $a \simeq 0.8$ when the MOG parameter reaches $\alpha = 1$. The constraints obtained from the observed jet power are represented by the red shaded regions, revealing notable differences between the cases of Lorentz factors $\Gamma = 2$ (left panel) and $\Gamma = 5$ (right panel). 
In the left panel, the allowed spin range begins at $a \simeq 0.26^{+0.08}_{-0.06}$ for the Kerr limit and increases to $a \simeq 0.72^{+0.16}_{-0.15}$ when $\alpha = 1$. 
In contrast, in the right panel corresponding to $\Gamma = 5$, the initial Kerr value is $a \simeq 0.571^{+0.123}_{-0.112}$, which rises to $a \simeq 1.289^{+0.12}_{-0.16}$ for $\alpha = 1$. A pronounced difference between the left and right panels emerges when simultaneously accounting for both observational constraints. Nevertheless, in both cases intersections between the allowed regions are present in  Fig. \ref{s4} for $\Gamma=2$ and $\Gamma=5$. In particular, the right panel demonstrates that the central solid curves corresponding to the radiative efficiency and jet power constraints intersect. This overlap enables the identification of parameter values that can simultaneously account for both the observed jet power and radiative efficiency of the source. Such a configuration is consistent with a system characterized by a relatively high Lorentz factor producing relativistic radio jets.

\subsection{Candidate GRO J1655-40}

\begin{figure*}
    \centering
    \includegraphics[width=0.5\linewidth]{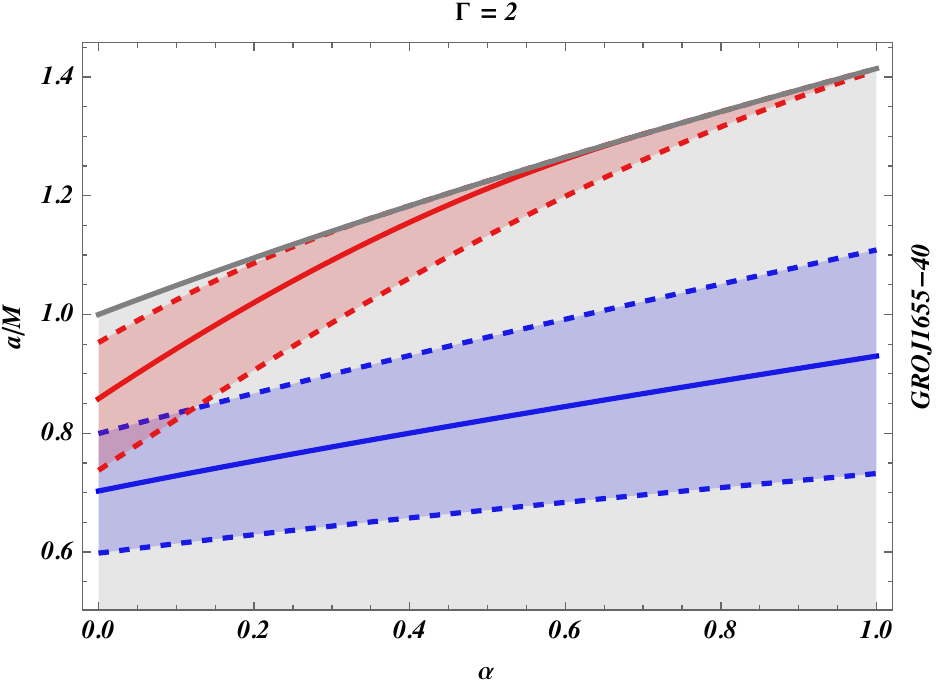}\includegraphics[width=0.5\linewidth]{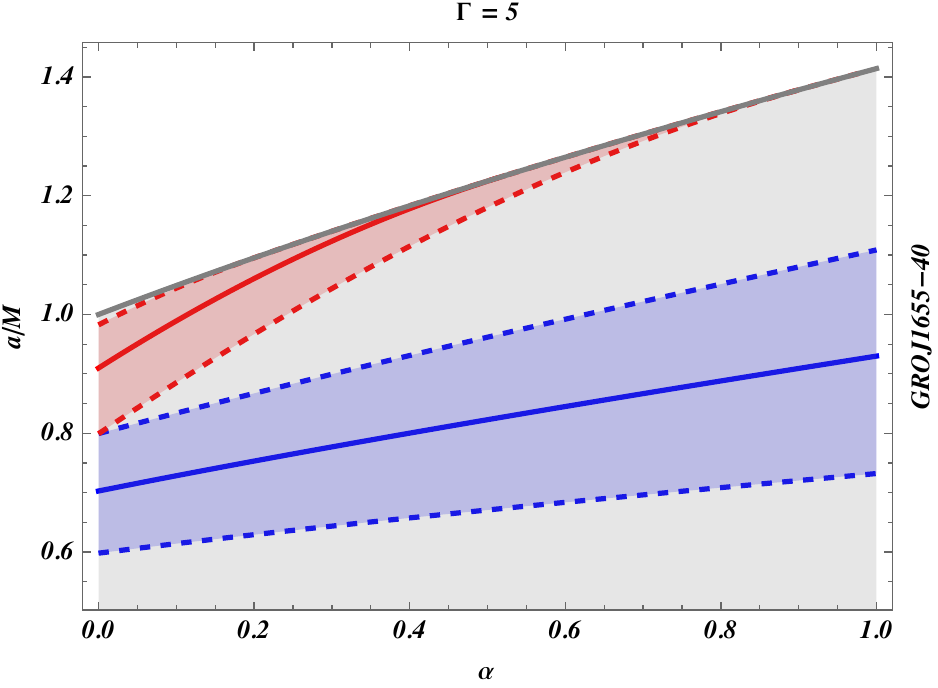}
    \caption{GRO J1655-40.The left panel corresponds to a Lorentz factor of $\Gamma = 2$, while the right panel shows $\Gamma = 5$. These plots illustrate that the rotating Kerr-MOG black hole can simultaneously account for the observed radiative efficiency, $\eta$ , and jet power, $P_{\text{jet}}$. Further details are discussed in the main text.
}
    \label{s5}
\end{figure*}

The spin value estimate obtained via the CFM gives  $a\simeq0.7\pm0.1$ at the 68\% CL \cite{51}.The blue shaded regions in Fig. \ref{s5} represent the constraints in the ($a,\alpha$) parameters plane inferred from the observed radiative efficiency, under the assumption that the Kerr-MOG black hole spacetime. Although the central values of the radiative efficiency and jet power constraints do not intersect, a partial overlap appears within the corresponding uncertainty regions. This indicates that the Kerr–MOG spacetime can account for the observations only when uncertainties are taken into consideration. Moreover, in the left panel corresponding to $\Gamma=2$, the lower bound of the jet power constraint intersects with the upper bound of the radiative efficiency constraint at approximately $\alpha=0.13$ for spin value of $a\approx0.85$. In contrast, the right panel shows almost no overlap between the jet power and radiative efficiency regions. Nevertheless, in both cases the allowed parameter ranges remain within the black hole region. A lower Lorentz factor appears to be more favorable for simultaneously reproducing both the radiative efficiency and jet power observations. From a theoretical perspective, by combining the radiative efficiency values listed in Table~\ref{table1} with the jet power measurements presented in Table~\ref{table2}, the overlapping regions of the corresponding constraints delineate the parameter space capable of satisfying the observational bounds, as shown in Fig.~\ref{s5}.

\subsection{Candidate GRS 1915+105}
\begin{figure*}[ht]
    \centering
    \includegraphics[width=0.5\linewidth]{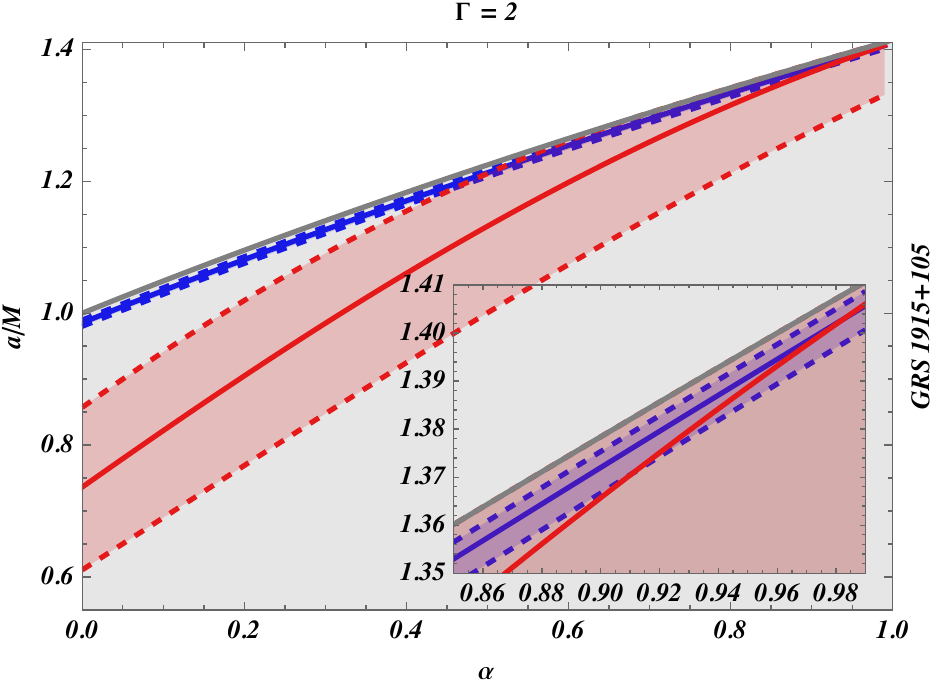}\includegraphics[width=0.5\linewidth]{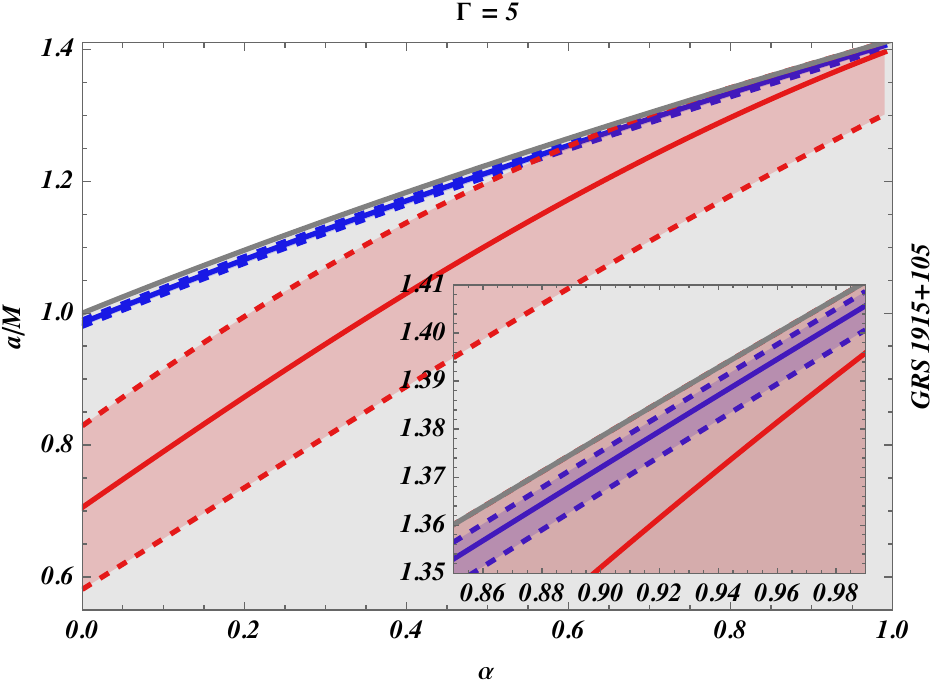}
    \caption{For GRS 1915+105, the left and right panels show that the parameter regions consistent with the observed radiative efficiency and jet power correspond to near-extremal spin values of the source across the full range of the MOG parameter $\alpha$ and spin $a$. Further details are provided in the main text. }
    \label{s6}
\end{figure*}

According to \cite{52}, the spin of GRS~1915+105 is constrained to be $a > 0.98$ at the 68\% CL. 
The blue shaded regions in Fig. \ref{s6} show the constraints on the parameters $a$ and $\alpha$ obtained from the observed radiative efficiency, assuming that the source is described by the Kerr--MOG spacetime. For this system, the blue dashed lines, which represent the uncertainty ranges, become very close to the central solid line as the MOG parameter increases, as illustrated in the zoomed-in panels. The regions corresponding to the observed jet power, shown in red, display similar behavior for both values of the Lorentz factor. 
It is also important to note that in the pure Kerr case, Eq. (\ref{jet}) is applicable only up to spin values of about $a \simeq 0.71$. For spin values larger than this limit, higher-order terms in \( \Omega_H \) should be considered \cite{43}. In our analysis, we include these terms in the jet power formula, $P_{BZ} = k \Phi^2_{\text{tot}} (\Omega_H^2 + \varepsilon \Omega_H^4 + \sigma \Omega_H^6)$, with  $\varepsilon \simeq 1.38$ and $\sigma \simeq -9.2$ taken from \cite{43}. However, we find that these extra terms have only a very small effect on the jet power and do not noticeably change the results. 
An inspection of both the left and right panels shows that the central values of the two observational constraints intersect only for the Lorentz factor $\Gamma=2$ at high spin value. This intersection occurs around $a\approx1.39$ and $\alpha\simeq0.98$. In the right panel, a narrow shaded overlap region is also present at high spin, although it arises from the uncertainty bounds indicated by the dashed lines rather than from the central values. For the upper Lorentz factor, the central curves do not intersect. This suggests that the observed radiative efficiency and jet power of the source (listed in Tables \ref{table1} and \ref{table2}) are only weakly consistent with each other for larger spacetime parameters in the rotating Kerr-MOG model.
 The Fig. \ref{s6} shows that the parameter region consistent with the observed radiative efficiency and jet powers correspond to near-extremal spin values of the source over the entire range of a free MOG parameter $\alpha$ and spin value $a$ as well.

\section{Conclusion}\label{sec5}

In this work, we have investigated the astrophysical implications of the rotating Kerr-MOG black hole spacetime by applying it to several well-known stellar-mass black hole candidates, namely A0620--00, H1743--322, XTE J1550--564, GRS1124--683, GRO J1655--40, and GRS1915+105. We have first introduced the basic properties of the Kerr--MOG geometry, which is characterized by the black hole mass, the spin parameter $a$, and the modified gravity parameter $\alpha$. Particular attention has been given to the structure of horizons and to the dependence of the ISCO on these spacetime parameters. Our analysis has shown that the MOG parameter significantly modifies the ISCO radius and therefore affects the radiative efficiency of accretion disks described within the Novikov-Thorne framework. In particular, we have demonstrated how the radiative efficiency depends simultaneously on the spin and the MOG parameter, leading to a degeneracy between these quantities when interpreting observational data. It has been shown that the dependence of the radiative efficiency on the parameter $\alpha$ is strongly influenced by the black hole spin. Specifically, for small values of the spin parameter, the radiative efficiency monotonically increases with $\alpha$, while for higher spin values it decreases as $\alpha$ increases. The effect of the parameter $\alpha$ on the efficiency-spin relation shows that larger values of $\alpha$ shift the curves toward higher efficiencies and larger spin values.

Next, we have used the observed radiative efficiencies of several X-ray binary systems to explore the possible values of the Kerr-MOG spacetime parameters. By comparing the theoretical predictions with the efficiencies inferred from the CFM, we have identified the regions in the $(a,\alpha)$ parameter space compatible with the observational data for each source. Our results indicate that an increase in the MOG parameter shifts the allowed spin values toward smaller magnitudes for sources with low or moderate spins, while for sources with higher spins it shifts the allowed spin values toward larger magnitudes in both cases increasing the interval of allowed spin values.

We have then extended our analysis to the relativistic jet power associated with transient jets observed in the same black hole systems. Using the Blandford-Znajek mechanism to estimate the jet power extracted from the rotating black hole, we have examined whether the Kerr-MOG spacetime can reproduce the observed jet energetics. The obtained results show that the jet power can indeed be explained within certain regions of the $(a,\alpha)$ parameter space for the considered sources. In general, the allowed spin range increases with the growth of the MOG parameter. The comparison has been performed for two representative values of the jet Lorentz factor, $\Gamma=2$ and $\Gamma=5$. The results show that the qualitative behavior of the allowed regions remains largely similar for most sources, although quantitative differences appear depending on the chosen Lorentz factor. In particular, considerable differences between the $\Gamma=2$ and $\Gamma=5$ cases are observed for the source GRS 1124-683, while small but noticeable differences appear for the source GRO J1655-40.

Finally, we have combined the constraints obtained from the radiative efficiency and the jet power in order to determine whether the Kerr-MOG spacetime can simultaneously reproduce both observational properties of the considered systems. Our analysis has shown that for several sources, including A0620-00, H1743-322, and XTE J1550-564, there exist overlapping regions in the parameter space where both observables can be consistently explained within the Kerr-MOG framework. For other objects, such as GRS1124-683 and GRO J1655-40, the compatibility between the two constraints becomes more sensitive to the assumed jet Lorentz factor, leading to partial or limited overlap regions. In the case of the highly spinning source GRS1915+105, the observational data tend to favor near-extremal spin configurations, and the corresponding allowed regions appear only in a narrow parameter range of the Kerr-MOG spacetime. This result is particularly surprising, since our previous studies (see e.g. \cite{narzilloev2024observed, Narzilloev2023h, Saidov:prd}) did not find any spacetime models capable of producing such overlapping regions for this source.

Overall, our results suggest that the Kerr-MOG black hole model can provide a viable framework for interpreting the observed radiative efficiency and jet power of several astrophysical black hole candidates. At the same time, the analysis highlights the importance of combining different observational probes in order to break the degeneracy between the spin and modified gravity parameters. Future observations with improved precision, together with larger samples of black hole systems, will make it possible to place stronger constraints on the MOG parameter and to further test the viability of modified gravity models in the strong-field regime.

\section{acknowledgement}
We acknowledge support by grant  FL-10425067111 from the Agency for Innovative Development of Uzbekistan.

%\boldsymbol{Data availability Statement}{The data that support the findings of this study are available from the corresponding author upon reasonable request.}

\bibliographystyle{apsrev4-1}
\bibliography{citations}
\end{document}